\documentclass[11pt,a4paper]{article}
\pdfoutput=1
\usepackage{jheppub}
\usepackage{multirow, graphicx,amssymb,url,mathrsfs,amsmath}
\usepackage{wrapfig,boxedminipage,epsfig}
\usepackage{amsxtra,amstext,latexsym,dsfont,amsfonts}
\usepackage{color}
\usepackage[dvipsnames]{xcolor}
\usepackage{float}
\usepackage{slashed}
\usepackage{calligra}
\DeclareFontShape{T1}{calligra}{m}{n}{<->s*[2.2]callig15}{}
\DeclareMathAlphabet{\mathcalligra}{T1}{calligra}{m}{n}
\usepackage{tikz} 
\usepackage[compat=1.1.0]{tikz-feynman}
\usepackage{feynmf}
\usepackage[title]{appendix}
\usepackage{ulem}

\renewcommand{\a}{\alpha}

\newcommand{\be}{\begin{equation}}
\newcommand{\ee}{\end{equation}}
\newcommand{\bea}{\begin{eqnarray}}
\newcommand{\eea}{\end{eqnarray}}

\newcommand{\dd}{\mathrm{d}}

\newcommand{\grad}{\nabla}

\title{\huge Deep learning bulk spacetime from boundary optical conductivity}

\author[a]{Byoungjoon Ahn,}
\author[b,c]{Hyun-Sik Jeong,}
\author[a,d]{Keun-Young Kim}
\author[a]{and Kwan Yun}

\affiliation[a]{Department of Physics and Photon Science, Gwangju Institute of Science and Technology, \\
123 Cheomdan-gwagiro, Gwangju 61005, Korea}
\affiliation[b]{Instituto de F\'isica Te\'orica UAM/CSIC, Calle Nicol\'as Cabrera 13-15, 28049 Madrid, Spain}
\affiliation[c]{Departamento de F\'isica Te\'orica, Universidad Aut{\'o}noma de Madrid, 28049 Madrid, Spain}
\affiliation[d]{Research Center for Photon Science Technology, Gwangju Institute of Science and Technology, \\
123 Cheomdan-gwagiro, Gwangju 61005, Korea}

\emailAdd{bjahn123@gist.ac.kr}
\emailAdd{hyunsik.jeong@csic.es}
\emailAdd{fortoe@gist.ac.kr}
\emailAdd{ludibriphy70@gm.gist.ac.kr}

\abstract{
We employ a deep learning method to deduce the \textit{bulk} spacetime from \textit{boundary} optical conductivity. We apply the neural ordinary differential equation technique, tailored for continuous functions such as the metric, to the typical class of holographic condensed matter models featuring broken translations: linear-axion models. We successfully extract the bulk metric from the boundary holographic optical conductivity. Furthermore, as an example for real material, we use experimental optical conductivity of $\text{UPd}_2\text{Al}_3$, a representative of heavy fermion metals in strongly correlated electron systems, and construct the corresponding bulk metric. To our knowledge, our work is the first illustration of deep learning bulk spacetime from \textit{boundary} holographic or experimental conductivity data.
}

\begin{document}
\maketitle

%%%%%%%%%%%%
%%%%%%%%%%%%
\section{Introduction}
The holographic correspondence, also known as holography or anti-de Sitter/conformal field theory (AdS/CFT) duality, is a valuable and widely used tool in various fields like quantum chromodynamics (QCD), condensed matter physics, hydrodynamics, and quantum information~\cite{Ammon:2015wua,Hartnoll:2016apf,Hartnoll:2009sz,Zaanen:2015oix,CasalderreySolana:2011us,Baggioli:2019rrs,Heller:2016gbp,Florkowski:2017olj,HBT2020,Chen:2021lnq}. It is often employed in a more practical, \textit{bottom-up design}, without resorting to its string theory origins~\cite{Maldacena:1997re} or issues associated with quantum gravity~\cite{Aharony:1999ti}. 
Conversely, holography is utilized as an effective framework for gaining insights into physical scenarios where conventional methods may not be applicable. It proves particularly advantageous, and sometimes even the sole viable approach, when dealing with systems exhibiting strong coupling, situations governed by complex many-body collective behavior lacking well-defined elementary excitations, and dissipative systems where formulating a proper finite temperature field theory is far from straightforward.

Considering the practical and application perspective of holography, it becomes evident that the primary challenge lies in the endeavor to closely align this framework with reality, thus ensuring its applicability to \textit{real-world physical situations}. An illustrative and quintessential example for this is the comparison between QCD, a non-abelian $SU(3)$ gauge theory, and $\mathcal{N}=4$ supersymmetric Yang-Mills theory in the large $N_c$ limit.\\

\paragraph{Condensed matter application and broken translations.}
In pursuit of practical real-world applications, another active area within applied holography is its utilization in the context of \textit{condensed matter systems}: dubbed as AdS/CMT. In the condensed matter, it is evident that Poincar\'e invariance is not maintained so that translations (and also rotations) are broken, often in a spontaneous manner, which is referred to as ``spontaneous" symmetry breaking instrumental in understanding the principles behind the rigidity and elasticity of matters. 

Furthermore, as in numerous situations (e.g., electronic transport), translations can also be broken ``explicitly", resulting in a finite DC conductivity (or a Drude peak in the optical conductivity) observed in common metals. 
Also, there are a mixed situation (e.g., a pinned charge density waves~\cite{Gruner:1988zz}) in which translations are broken both explicitly and spontaneously, known as the pseudo-spontaneous limit, leading to a finite frequency peak in optical conductivity.

As such, in order to describe more realistic situations for the condensed matter application in holography, it is imperative to develop and understand holographic gravitational models that encompass the broken translational invariance in the dual boundary theory. Historically, this endeavor initiated with the development of holographic models incorporating bulk fields featuring spatially dependent boundary conditions, which are designed to resemble an explicit lattice source. 

Depending on the approach employed to break translational symmetry, holographic models can be categorized into two classes: homogeneous and inhomogeneous models. In this context, homogeneous signifies that the spacetime geometry relies exclusively on the holographic bulk direction, maintaining independence from field theory directions. Otherwise, it comes to inhomogeneous models.

As an example, homogeneous model includes helical lattice~\cite{Donos:2012js, Donos:2014oha, Donos:2014gya}, Q-lattice~\cite{Donos:2013eha, Donos:2014uba}, and linear-axion model~\cite{Andrade:2013gsa, Baggioli:2021xuv}. In the case of the linear-axion model, translations are broken by the massless scalar fields that vary linearly with spatial direction.\footnote{The linear-axion model is closely associated with the St\"uckelburg formulation of a massive gravity theory~\cite{Vegh:2013sk, Davison:2013jba, Blake:2013bqa, Blake:2013owa, Alberte:2015isw}.} On the other hand, inhomogeneous models are constructed by imposing periodic boundary conditions either on a scalar field (i.e., scalar lattice) or on the chemical potential (i.e., ionic lattice). Pioneering contributions to this field can be found in~\cite{Horowitz:2012ky, Horowitz:2012gs,Nakamura:2009tf,Donos:2011bh}.

It is worth noting that the homogeneous model can be technically more tractable than inhomogeneous model in the sense that one needs to solve partial differential equations for inhomogeneous models due to the periodic boundary conditions, while in the homogenous cases, it is reduced to solving ordinary differential equations.

\paragraph{Holographic linear-axion models.}
Furthermore, homogeneous model, especially \textit{linear-axion models}~\cite{Andrade:2013gsa, Baggioli:2021xuv}, can be more appealing because of a closed-form analytical background as well as its feasibility to consider all types of translational symmetry breaking patterns from explicit to spontaneous one.\footnote{The symmetry breaking pattern in linear-axion model can be controlled by the change of the asymptotic boundary condition of the scalar field~\cite{Baggioli:2021xuv}. See also \cite{Ahn:2022azl,Jeong:2023las} for the similar analysis with the vector fields in order to study the dynamical electromagnetism in the dual boundary field theory.} Its novel and celebrated results have been achieved from the analysis of conductivity~\cite{Davison:2013txa,Gouteraux:2014hca,Blauvelt:2017koq,Alberte:2017cch,Jeong:2018tua,PhysRevLett.120.171602,Ammon:2019wci,Ahn:2019lrh,Jeong:2021wiu,Baggioli:2022pyb,Ahn:2023ciq} (such as anomalous properties of strange metals, high-Tc superconductivity, and pinning structure in optical conductivity), fermionic spectral functions~\cite{Jeong:2019zab}, transport coefficients such as diffusivity~\cite{Davison:2014lua,Blake:2016jnn,Blake:2017qgd,Baggioli:2017ojd,Ahn:2017kvc,Davison:2018ofp,Blake:2018leo,Arean:2020eus,Liu:2021qmt,Jeong:2021zhz,Wu:2021mkk,Jeong:2021zsv,Huh:2021ppg,Baggioli:2022uqb,Jeong:2023ynk}\footnote{In recent years, a novel phenomenon known as pole-skipping has been employed to establish a connection between thermal diffusive dynamics and quantum chaos~\cite{Grozdanov:2017ajz,Blake:2017ris,Blake:2018leo,Grozdanov:2019uhi,Blake:2019otz,Natsuume:2019xcy,Natsuume:2019sfp,Natsuume:2019vcv,Ceplak:2019ymw,Ahn:2019rnq,Ahn:2020bks,Abbasi:2020ykq,Liu:2020yaf,Ramirez:2020qer,Ahn:2020baf,Natsuume:2020snz,Ceplak:2021efc,Jeong:2021zhz,Natsuume:2021fhn,Blake:2021hjj,Jeong:2022luo,Wang:2022mcq,Amano:2022mlu,Yuan:2023tft,Grozdanov:2023txs,Natsuume:2023lzy,Ning:2023pmz,Grozdanov:2023tag,Jeong:2023rck,Natsuume:2023hsz,Abbasi:2023myj,Jeong:2023ynk}. Leveraging broken translations, pole-skipping offers a new perspective for comprehending the diffusive process within the framework of quantum chaos.}, and collective dynamics of strongly coupled phase.\footnote{The holographic axion model is not just a makeshift approach for breaking translational symmetries; rather, its framework can be consistently related to and derived from the conventional effective field theory formulations, as elaborated in recent review~\cite{Baggioli:2022pyb}.} See also \cite{RezaMohammadiMozaffar:2016lbo,Yekta:2020wup,Li:2019rpp,Zhou:2019xzc,Huang:2019zph,Jeong:2022zea,HosseiniMansoori:2022hok} for the quantum information applications. For a comprehensive and recent review of the holographic axion model and an extensive list of references, we refer \cite{Baggioli:2021xuv}.

\paragraph{Neural ordinary differential equations and holography.}
In essence, the linear-axion model has served as a pivotal toy gravitational model for studying strongly coupled condensed matter systems. In particular, it operates as a bottom-up approach where gravitational bulk physics is implemented to depict the realistic dual boundary condensed matter systems characterized by broken translational symmetries.

In this paper, employing state-of-the-art techniques in applied holography, specifically \textit{neural ordinary differential equations (neural ODE)}~\cite{Chen:2018aa} in machine learning, we explore the holographic condensed matter application in the opposite direction.
In other words, we investigate the geometric bulk physics (such as the bulk spacetime metric) from the boundary condensed matter physics, i.e. it is the inverse problem.
For this purpose, we focus on the original linear-axion model~\cite{Andrade:2013gsa} where translations are broken explicitly. Subsequently, we utilize the \textit{optical conductivity} as the input data from the dual boundary perspective.

Machine learning stands out as a significant tool in both theoretical and experimental physics, proving especially powerful in domains characterized by extensive data. However, its application in the more formal realms of theoretical physics remains somewhat restricted.

When approaching the AdS/CFT study as a problem of determining the bulk theory for a given boundary quantum field theory (QFT) data, it may take on the nature of data science. This involves the extraction of features from the extensive QFT data to interpret it as a higher-dimensional gravity theory. In addressing such problems, machine learning proves instrumental in identifying the underlying bulk theory, i.e., an illustrative example of reconstructing the bulk geometry.

Efficient holographic modeling has been demonstrated through the utilization of machine learning techniques \cite{Hashimoto:2018ftp,You:2017guh,Hashimoto:2018bnb,Hu:2019nea,Hashimoto:2019bih,Han:2019wue,Tan:2019czc,Yan:2020wcd,Akutagawa:2020yeo,Hashimoto:2020jug,Chen:2020dxg,Song:2020agw,Hashimoto:2021ihd,Lam:2021ugb,Park:2022fqy,Katsube:2022ofz,Hashimoto:2022eij,Li:2022zjc,Jejjala:2023aa,Park:2023slm}. Notably, a method proposed in \cite{Hashimoto:2018ftp,You:2017guh} treats the ``discretized" bulk geometry as a neural network, with the network weights representing the bulk spacetime metric. The input data for the neural network comprises the information from the boundary QFT. Consequently, deep learning (DL), a form of machine learning employing deep neural networks, exhibits similarities with the AdS/CFT correspondence and functions as a solver for the inverse problem. Upon completing the training of the neural network, the bulk metric can be determined. This is called the AdS/DL correspondence.\footnote{Machine learning techniques are commonly employed in holography through two main methodologies: AdS/DL and EFL (Entanglement Feature Engineering). In the case of EFL, the approach entails the construction of a tensor network within a given Anti-de Sitter space, and Boltzmann machines are employed to optimize the tensor network~\cite{You:2017guh,Hashimoto:2019bih,Lam:2021ugb}.} {For those who want to understand the essential idea of AdS/DL in a simpler setup easily, we refer to \cite{Song:2020agw}, where the idea was realized in a classical mechanics problem.}

In this manuscript, we investigate the scenario with finite momentum relaxation to assess the efficacy of machine-learning holographic CMT in more realistic situations. Our objective is to deduce the bulk spacetime metric from the electric optical conductivity data in the presence of momentum relaxation (i.e., broken translations). 

In particular, we choose the neural ODE~\cite{Chen:2018aa} as the solver, which is a machine learning technique suitable for ``continuous" systems as opposed to discretized ones. It is noteworthy that employing the neural ODE approach, which replaces the weights in neural networks with continuous functions, not only enhances accuracy but also provides a natural interpretation of the metric function~\cite{Hashimoto:2020jug}.

It is pertinent to highlight that the AdS/DL correspondence in this context has found applications in holography. Examples include the magnetization curve of the strong correlated materials~\cite{Hashimoto:2018ftp}, a lattice QCD data of the chiral condensate at a finite temperature~\cite{Hashimoto:2018bnb, Hashimoto:2020jug}, hadron spectra~\cite{Akutagawa:2020yeo}, the dilaton potential in improved holographic QCD  from the experimental data of the $\rho$ meson spectrum~\cite{Hashimoto:2021ihd,Hashimoto:2022eij}, the complex frequency-dependent shear viscosity in strongly coupled systems in AdS/CMT~\cite{Yan:2020wcd}, optical conductivity with the translation invariance~\cite{ Li:2022zjc}, entanglement entorpy~\cite{Park:2022fqy,Park:2023slm}.\footnote{References of significance can also be found without holography. These references, for instance, investigate the methodologies of machine learning within condensed matter physics~\cite{Carrasquilla_2017, Carrasquilla:2020mas, Bedolla-Montiel:2020rio, Suwa:2018twu}, as well as in the fields of nuclear physics or QCD~\cite{Boehnlein:2021eym,Chen:2021giw, Zhou:2023pti}.}

{Note that the authors in \cite{Li:2022zjc} has also been deduced the bulk metric from conductivity using deep learning in holography. However, in \cite{Li:2022zjc}, the so called reduced conductivity, which is not exclusively defined at the boundary but extends across the entire bulk dimension, was used. We plan to improve this work by using the conductivity strictly defined at the boundary.}

It is also worth noting that, in addition to the advantage of dealing with a realistic setup involving finite momentum relaxation, our analysis introduces a further technical development compared to the previously mentioned works. In prior research, attempts were made to construct neural networks based on the bulk equations of motion of a single field (scalar, metric tensor, or gauge field), namely, the decoupled equations. However, our setup, which incorporates finite momentum relaxation and density, presents a challenge as it involves coupled equations of motion encompassing all scalar, metric, and gauge fields.

This paper is organized as follows. 
In section \ref{sec2}, we provide a review of the linear-axion model and holographic calculations pertaining to the  electric optical conductivity.
In section \ref{sec3}, we introduce the neural ODE with a  metric ansatz tailored to the holographic CMT discussed in section \ref{sec2}. We utilize both the holographic conductivity and the real materials as input data of neural ODE.
Section \ref{sec4} is devoted to conclusions.

%%%%%%%%%%%%
%%%%%%%%%%%%
\section{Holographic model with momentum relaxation}\label{sec2}
In this section, we examine the standard holographic setup employed for investigating electric optical conductivity in the presence of momentum relaxation. As outlined in the introduction, we introduce the linear-axion model~\cite{Andrade:2013gsa, Baggioli:2021xuv} for this purpose. It is instructive to note that the machine learning procedure in the following section is grounded in this holographic setup, where the optical conductivity serves as the input data.

%%%%%%%%%%%%
\subsection{Linear-axion model: action and ansatz}
The linear-axion model corresponds to a specific example of the Einstein-Maxwell-Axion model. In 4-dimensional spacetime, its action is given as
\begin{align}\label{action}
\begin{split}
    \mathcal{S} = \int \dd^4x \sqrt{-g} \left( R + 6 - \frac{1}{4} F_{ab}F^{ab} - \frac{1}{2}\sum_{I=1}^{2} (\partial X_I)^2 \right) \,,
\end{split}
\end{align}
where we have chosen units for simplicity where the gravitational constant $16\pi G=1$, and the AdS length $L=1$. The linear-axion model given by \eqref{action} incorporates two matter fields: the U(1) gauge field $A$ with its associated field strength $F=\dd A$ and the axion field $X$. From the perspective of the boundary field theory, the former serves to manifest finite density or chemical potential, while the latter is introduced to break translational invariance, leading to momentum relaxation.

The action \eqref{action} produces the equations of motion 
\begin{align}\label{field_eq}
\begin{split}
& R_{ab} - \frac{1}{2} g_{ab} \left( R + 6 - \frac{1}{4} F_{ab}F^{ab} - \frac{1}{2}\sum_{I=1}^{2} (\partial X_I)^2 \right) - F_{ac} F_b^{c} - \frac{1}{2}\sum_{I=1}^{2} \partial_a X_I \partial_b X_I = 0 \,, \\
& \grad^a F_{ab} = 0 \,,\qquad \grad_a \grad^a X_I = 0 \,,
\end{split}
\end{align}
and for our analysis, we employ the following ansatz:
\begin{align}\label{metric}
\begin{split}
\dd s^2 &= \frac{1}{z^2} \left[ -f(z) \dd t^2 +\frac{\dd z^2}{f(z)} + \dd x^2 + \dd y^2 \right] \,, \\
A &= \mu \left(1 - z\right) \dd t \,, \qquad X_1 = \a \, x \,, \quad X_2 = \a \, y \,,
\end{split}
\end{align}
where $\mu$ represents the chemical potential, and $\alpha$ denotes the strength of momentum relaxation. Here, for numerical convenience in machine learning, we fix the horizon at $z_H=1$ where $f(z_H)=0$.

Plugging \eqref{metric} into the equations \eqref{field_eq}, one can find the analytic solution for the emblackening factor $f(z)$ as
\begin{align}\label{true_metric}
\begin{split}
 f(z) = 1 - \frac{\alpha^2}{2}z^2 -\left( 1 - \frac{\alpha^2}{2} + \frac{\mu^2}{4} \right) z^3  + \frac{\mu^2}{4}z^4 \,,
\end{split}
\end{align}
then the Hawking temperature ($T_H$) can be read as
\begin{align}\label{tem}
\begin{split}
T_H := - \frac{f'(1)}{4\pi} = \frac{12 - 2 \alpha^2 - \mu^2}{16 \pi} \,.
\end{split}
\end{align}
%

%%%%%%%%%%%%
\subsection{Holographic optical conductivity}
Next, we review the holographic optical conductivity following the approach presented in~\cite{Andrade:2013gsa}. To do so, we consider the bulk fluctuations based on the background geometry \eqref{metric}:
\begin{align}\label{}
\begin{split}
\delta g_{tx} = e^{-i \omega t} \frac{h_{tx}(z)}{z^2} \,, \qquad \delta A_{x} = e^{-i \omega t} a_{x}(z) \,, \qquad \delta X_1 = e^{-i \omega t} \frac{\psi_x(z)}{\alpha} \,,
\end{split}
\end{align}
where $\omega$ represents the frequency, and the remaining perturbations are decoupled.

In addition, it is also convenient \cite{Andrade:2013gsa} to introduce
\begin{equation}\label{phi_tranf}
\phi(z) := -\frac{f(z) \psi_x'(z)}{\omega \, z } \,.
\end{equation}
Subsequently, the equations of motion for the fluctuations can be expressed as
\begin{align}\label{EOMFLUC}
\begin{split}
    &a_x''(z) + \frac{f'(z)}{f(z)} a_x'(z) +  \left( \frac{\omega^2}{f(z)^2} - \frac{\mu^2 z^2}{f(z)} \right) a_x(z) - \frac{i\mu z}{f(z)} \phi(z) = 0 \,, \\
    &\phi''(z) +  \frac{f'(z)}{f(z)}  \phi'(z) + \left( \frac{\omega^2}{f(z)^2} - \frac{\alpha^2}{f(z)} - \frac{f'(z)}{zf(z)}  \right) \phi(z) + \frac{i\alpha^2\mu z}{f(z)} a_x(z) = 0  \,,
\end{split}
\end{align}
where the equation of $h_{tx}$ is algebraically eliminated.
Note that using \eqref{EOMFLUC} one can find that the fluctuations behave near AdS boundary ($z\rightarrow0$) as
\begin{align}\label{BCSET}
\begin{split}
a_x  = a_{x}^{(S)} + a_{x}^{(R)} z + \cdots \,, \qquad \phi = \phi^{(S)} + \phi^{(R)} z + \cdots \,,
\end{split}
\end{align}
where the leading coefficients $\left(a_{x}^{(S)},\,\phi^{(S)}\right)$ denote the source, while the sub-leading coefficients $\left(a_{x}^{(R)},\,\phi^{(R)}\right)$ are interpreted as the response by the holographic dictionary.\\

The electric optical conductivity, denoted as $\sigma(\omega)$, can be derived using the Kubo formula in terms of the boundary coefficients presented in \eqref{BCSET}:
\begin{align}\label{conductivity}
\begin{split}
   \sigma(\omega) =  \frac{1}{i \omega} G^{R}_{j^xj^x}(\omega) = \frac{1}{i \omega} \frac{a_x^{(R)}}{a_x^{(S)}} \,,
\end{split}
\end{align}
where $G^{R}_{j^xj^x}$ is the current-current retarded Green's function. 

Four remarks regarding \eqref{conductivity} are worth noting.
First, the second equality in \eqref{conductivity} holds when $a_x^{(S)}$ is the only non-zero source~\cite{Andrade:2013gsa}. To do this, one needs to impose the sourceless condition for the axion field, i.e., $\phi^{(S)}=0$.

Second, when solving the fluctuation equations \eqref{EOMFLUC} and determining the coefficients \eqref{BCSET} for evaluating \eqref{conductivity}, all the fluctuations fields satisfy ingoing boundary conditions at the horizon, as specified in \eqref{FOCOEFF}.

Third, in the DC limit ($\omega\rightarrow0$), the analytic expression of the electric conductivity can be given by
\begin{align}\label{}
\begin{split}
   \sigma(\omega\rightarrow0) =  1 + \frac{\mu^2}{\alpha^2} \,,
\end{split}
\end{align}
where one can find that the DC conductivity has a finite value, consistent with the physics of momentum dissipation ($\alpha\neq0$).

Last but not least, to utilize the holographic electric conductivity \eqref{conductivity} in the subsequent machine learning analysis, we consistently enforce the three aforementioned conditions throughout this paper.

%%%%%%%%%%%%
%%%%%%%%%%%%
\section{Machine learning and holographic CMT}\label{sec3}
The AdS/DL correspondence represents an emerging research direction within the field of utilizing machine learning methodologies for addressing physics challenges~\cite{Hashimoto:2018bnb,Hashimoto:2018ftp,Hashimoto:2019bih,Akutagawa:2020yeo,Hashimoto:2020jug,Yan:2020wcd,Hashimoto:2021ihd,Hashimoto:2022eij,Li:2022zjc}. This paradigm introduces artificial intelligence for the elucidation of the holographic bulk theory underlying quantum systems on the holographic boundary.

Enforcing the holographic principle~\cite{Maldacena:1997re,Gubser:1998bc,Witten:1998qj} within a deep neural network framework, the AdS/DL correspondence posits the neural network as the classical equation of motion governing the evolution of fields on a \textit{discretized} curved spacetime.\footnote{Notably, advancements have been achieved through the neural network renormalization group (neural RG)~\cite{Hu:2019nea}, which systematically constructs the precise holographic mapping between boundary and bulk field theories at the partition function level.}

The central tenet of this framework asserts that the emergent dimensionality of the holographic bulk aligns with the depth dimension of the deep neural network, wherein the neural network itself serves as the representation of the bulk spacetime. As the neural network assimilates the holographic boundary data from its input layer, the optimization of network weights in deeper layers ensues, culminating in an optimal holographic bulk description corresponding to the given boundary data.

%%%%%%%%%%%%
\subsection{Neural ordinary differential equations}
Nevertheless, the current progress has predominantly relied on discretizing the holographic bulk dimension due to the inherent discreteness of neural network layers in a deep learning approach. For instance, one can solve the ordinary differential equation (ODE) within the Residual Network~\cite{E:2017} by employing the Euler method.

It is informative to revisit the Euler method for later use, which is a numerical technique for solving ODE with a given initial value. Let us consider a first-order ODE of the form:
\begin{align}
\begin{split}
\partial_{z} \mathcal{F} = \mathcal{G}\left(z, \mathcal{F}\right) \,,
\end{split}
\end{align}
where $\mathcal{F}$ is the unknown function of $z$, and $\mathcal{G}\left(z, \mathcal{F}\right)$ is some known function. The goal is to find an approximation to $\mathcal{F}(z)$ over a specific range of $z$.

The Euler method works by discretizing the $z$ interval into small steps. Given an initial value $\mathcal{F}_1$ at $z_1$, the next value $\mathcal{F}_2$ is estimated using the formula:
\begin{align}\label{EM1}
\begin{split}
& \mathcal{F}_{2} = \mathcal{F}_{1} + h\left(\mathcal{F}_{1}; z_1 \right) \cdot \Delta z  \,, \quad\quad  h\left(\mathcal{F}_{n}; z_n \right) =  \partial_{z} \mathcal{F} \big|_{z=z_n}  \,, 
\end{split}
\end{align}
where $\Delta z$ is the step size, representing the width of each interval. The next value can also be found straightforwardly as
\begin{align}\label{EM2}
\begin{split}
\mathcal{F}_{3} \,=\, \mathcal{F}_{2} \,+\, h\left(\mathcal{F}_{2}; z_2 \right) \cdot \Delta z \,=\, \mathcal{F}_{1} \,+\, \left[h\left(\mathcal{F}_{1}; z_1 \right) \,+\, h\left(\mathcal{F}_{2}; z_2 \right)\right] \cdot \Delta z \,,
\end{split}
\end{align}
where we used \eqref{EM1} in the second equality.
Then, the general recursive formula for the Euler method can be expressed as 
\begin{align}\label{EM3}
\begin{split}
 \mathcal{F}_{N} \,=\, \mathcal{F}_{1} \,+\, \sum_{n=1}^{N-1} \, h\left(\mathcal{F}_{n}; z_n, \,\theta_n\right) \cdot \Delta z \,.
\end{split}
\end{align}
In this way, one can find the approximate numerical solution $\mathcal{F}(z)$. Note that we also include $\theta_n$ in \eqref{EM3} to express a general hidden function in ODE, which will be the trainable parameters in neural network. The utilization of the AdS/DL correspondence with the Residual Network is explicated in the literature that includes \cite{Hashimoto:2018ftp, Hashimoto:2018bnb, Li:2022zjc} where the metric function $f_n(z_n)$ serves as the training variable.

It is also beneficial to note that the Euler method can accumulate errors, especially over long intervals or when dealing with complicated ODEs. There are more sophisticated numerical methods, such as the Runge-Kutta methods, which offer improved accuracy by using multiple function evaluations within each step. In this paper, we employ the Runge-Kutta method for this purpose: the detailed information can be found in Appendix \ref{AP1}.

\paragraph{Neural ordinary differential equation.}
It is desirable to establish continuity in the above approach, as a holographic spacetime is essentially smooth. To achieve this objective, leveraging the recent advancements in the neural ordinary differential equation (neural ODE) approach becomes instrumental~\cite{Chen:2018aa}. It is a \textit{continuous} version of Residual Network~\cite{Sander:2022aa}, namely
\begin{align}\label{EM4}
\begin{split}
 \mathcal{F}_{fin} \,=\, \mathcal{F}_{ini} \,+\, \int_{z_{ini}}^{z_{fin}} \, h\left(\mathcal{F}; z, \,\theta\right) \dd z \,, \quad\quad  h\left(\mathcal{F}; z, \, \theta \right) =  \partial_{z} \mathcal{F} \,. 
\end{split}
\end{align}
Notice that the subscript $n$ in \eqref{EM3} has been omitted\footnote{Note that, when we implement numerical analysis, the integral is still a discrete sum. However, the integral notation of \eqref{EM4} emphasizes that we no longer need to discretize for ourselves if we employ the adaptive ODE solver. See also  Appendix \ref{AP2}.}, and the training variables, denoted as $\theta$, can be continuous, e.g., the metric function $f(z)$. 

Substituting the discrete neural network with the neural ODE not only provides a natural interpretation of the metric function in a continuous spacetime but also significantly enhances accuracy. Notably, in prior studies, such as~\cite{Hashimoto:2018bnb, Hashimoto:2018ftp, Akutagawa:2020yeo, Yan:2020wcd}, the discrete nature of the neural network necessitated the introduction of regularization terms to reduce discretization artifacts and ensure the smoothness of network weights. This regularization is unnecessary in the neural ODE approach.

The application of the AdS/DL correspondence with the neural ODE is elucidated within the holographic QCD framework~\cite{Hashimoto:2020jug}. In this context, the neural ODE is employed to find the emergence of bulk spacetime from provided data on the chiral condensate of lattice QCD. In this manuscript, we utilize the neural ODE approach within the context of holographic CMT, with the provided dataset consisting of optical electric conductivity.

\paragraph{The neural ODE within holographic CMT.}
Next, we provide the procedure for employing the neural ODE within the framework of our holographic CMT investigation. To begin with, our bulk equations of motion \eqref{EOMFLUC} can be translated into the neural ODE equation through the following identifications:
\begin{align}
\begin{split}
\mathcal{F}(z) \,\,\longleftrightarrow\,\, a_x(z) \,\,\,\text{and}\,\,\, \phi(z)\,, \qquad\quad  \theta \,\,\longleftrightarrow\,\, f(z) \,.
\end{split}
\end{align}
Here, the bulk metric function $f(z)$ is equivalent to the neural network weight $\theta$. To ensure a finite network depth, we incorporate the IR and UV cutoffs for the metic as 
\begin{align}
\begin{split}\label{cutoff}
z_{ini}=0.9999 \,, \qquad z_{fin}=0.0001.
\end{split}
\end{align}
\begin{figure}[]
\centering
\includegraphics[width=\textwidth]{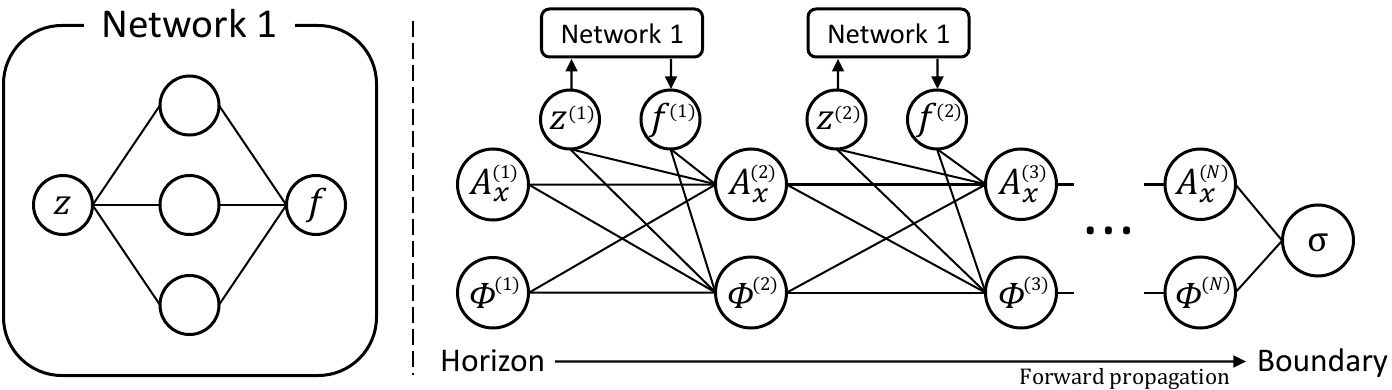}
\caption{A schematic picture for the architecture of the neural networks. This comprises two types of network: the initial one being the deep neural network mapping from $z$ to $f$, identified as ``Network 1". The second network is dedicated to the propagating bulk fields across $N$ layers, spanning from the horizon to the boundary.}\label{SPANODE}
\end{figure}

It is convenient to define the our bulk fields $\left(a_x,\, \phi \right)$ as 
\begin{align}\label{RDBF}
\begin{split}
A_{x}(z) := (1-z)^{-\frac{i\omega}{f'(1)}} a_x(z) \,, \qquad 
\Phi(z) := (1-z)^{-\frac{i\omega}{f'(1)}} \phi(z)  \,,
\end{split}
\end{align}
to ensure the incoming boundary condition at the black hole horizon. Then, the original bulk equations \eqref{EOMFLUC} can be rewritten 
\begin{align}\label{RDEOM}
\begin{split}
\partial_z^2 A_{x} &\,=\, \zeta \, \partial_z A_{x} \,+\, \left(\frac{z^2\mu^2}{f} \,-\, \xi\right)A_{x} \,+\, \frac{iz\mu}{f}\Phi  \,, \\
\partial_z^2 \Phi &\,=\, \zeta \, \partial_z \Phi \,+\, \left(\frac{\alpha^2}{f} + \frac{f'}{z f} - \xi\right)\Phi \,-\, \frac{iz\alpha^2\mu}{f}A_{x} \,,
\end{split}
\end{align}
where
\begin{align}
\begin{split}
\zeta := \frac{2i\omega}{(1-z)f'(1)} - \frac{f'(z)}{f(z)} \,, \quad\,\,  \xi := \frac{\omega^2}{f(z)^2} + \frac{i\omega}{(1-z)f'(1)}\left(\frac{i\omega}{(1-z)f'(1)} - \frac{1}{1-z} - \frac{f'(z)}{f(z)}\right) \,.
\end{split}
\end{align}
For the later use, let us also denote the first-order derivative of the redefined bulk fields \eqref{RDBF} at the horizon
\begin{align}\label{FOCOEFF}
\begin{split}
\partial_z A_{x}(1) &\,=\, -\left(\frac{i\omega f''(1)}{2f'(1)^2} + \frac{i\mu^2}{2\omega-if'(1)}\right) A_{x}(1) \,+\, \frac{\mu}{2\omega - if'(1)} \Phi(1)  \,, \\
\partial_z \Phi(1) &\,=\, -\frac{\alpha^2\mu}{2\omega-if'(1)}A_{x}(1) \,-\, \left(\frac{i\omega f''(1)}{2f'(1)^2} + \frac{i\alpha^2 + if'(1)}{2\omega - if'(1)}\right) \Phi(1) \,,
\end{split}
\end{align}
where these are obtained from the horizon expansion of equations \eqref{RDEOM} together with the horizon condition $f(1)=0$. They are simply saying that the sub-leading coefficients of the fields near the horizon, $\left(\partial_z A_{x}(1)\,, \partial_z\Phi(1)\right)$, are determined by the leading coefficients $\left(A_{x}(1)\,,\Phi(1)\right)$.

Now, we present the algorithm for solving equations \eqref{RDEOM} using the neural network. For illustrative purposes, we initially employ the discretized version, i.e., \eqref{EM3} for explanatory clarity. Our algorithm begins from the horizon and progresses towards the AdS boundary across $N$ layers. Hereafter, we adopt shorthand notation for the bulk fields assessed at each bulk dimension $z$ (or layers), denoted as $\left(A_{x}^{(1)},\,\Phi^{(1)}\right)$ for the fields evaluated at the horizon and $\left(A_{x}^{(N)},\,\Phi^{(N)}\right)$ for those at the $N$-th layer.\\

The algorithm is given as follows.
\begin{enumerate}
\item{$\left(A_{x}^{(1)},\,\Phi^{(1)}\right)$, \textbf{\textit{the fields at the 1st layer}}, are assigned by random values at the specified parameters of interest: $\left(\omega,\, \mu,\, \alpha \right)$.}
%%%
\item{Find $\left(\partial_zA_{x}^{(1)},\, \partial_z \Phi^{(1)}\right)$ by solving \eqref{FOCOEFF}. Also using the bulk equations \eqref{RDEOM}, identify the second-derivative values: $\left(\partial_z^2 A_{x}^{(1)},\, \partial_z^2 \Phi^{(1)}\right)$.}
%%%
\item{$\left(A_{x}^{(2)},\,\Phi^{(2)}\right)$, \textbf{\textit{the fields at the 2nd layer}}, can be obtained by employing \eqref{EM3}, i.e., $A_x^{(2)} = A_x^{(1)} + \partial_z A_x^{(1)} \, \Delta z$ \,and\, $\Phi^{(2)} = \Phi^{(1)} + \partial_z \Phi^{(1)} \, \Delta z$.}
%%%
\item{Find $\left(\partial_zA_{x}^{(2)},\, \partial_z \Phi^{(2)}\right)$ by employing \eqref{EM3}, i.e., $\partial_z A_x^{(2)} = \partial_z A_x^{(1)} + \partial_z^{2} A_x^{(1)} \, \Delta z$ \,and\, $\partial_z \Phi^{(2)} = \partial_z \Phi^{(1)} \,+\, \partial_z^{2} \Phi^{(1)} \, \Delta z$. Also find the second-derivative values, $\left(\partial_z^2 A_{x}^{(2)},\, \partial_z^2 \Phi^{(2)}\right)$, by \eqref{RDEOM}.}
%%%
\item{For $N\geq3$, repeat step 3 and 4, in other words,\\

$\left(A_{x}^{(N)},\,\Phi^{(N)}\right)$, \textbf{\textit{the fields at the $N$-th layer}}, can be obtained by employing \eqref{EM3}, i.e., $A_x^{(N)} = A_x^{(N-1)} + \partial_z A_x^{(N-1)} \, \Delta z$ \,and\, $\Phi^{(N)} = \Phi^{(N-1)} + \partial_z \Phi^{(N-1)} \, \Delta z$.\\

Find $\left(\partial_zA_{x}^{(N)},\, \partial_z \Phi^{(N)}\right)$ by employing \eqref{EM3}, i.e., $\partial_z A_x^{(N)} = \partial_z A_x^{(N-1)} + \partial_z^{2} A_x^{(N-1)} \, \Delta z$ \,and\, $\partial_z \Phi^{(N)} = \partial_z \Phi^{(N-1)} \,+\, \partial_z^{2} \Phi^{(N-1)} \, \Delta z$. Also find the second-derivative values, $\left(\partial_z^2 A_{x}^{(N)},\, \partial_z^2 \Phi^{(N)}\right)$, by \eqref{RDEOM}.}
\end{enumerate}
The illustrative architecture depicting the algorithm outlined above is presented in Fig. \ref{SPANODE}.

Before continuing, one remark is in order. It is imperative to elucidate the procedure for determining the bulk metric $f_n(z_n)$ at each layer, denoted as $f_n^{(N)}(z_n)$, in the algorithm.
In the context of discrete neural networks, as discussed in works~\cite{Hashimoto:2018bnb, Hashimoto:2018ftp, Akutagawa:2020yeo, Yan:2020wcd}, the random values are introduced for $f_n^{(N)}$. Consequently, the metric function $f_n$ is generally non-smooth and exhibits a seemingly random pattern. As a result, regularization terms become imperative to mitigate discretization artifacts, ensuring the smoothness of network weights. It is noteworthy, however, that such regularization is unnecessary in the neural ODE approach.

For the case of neural ODE, which represents the continuous counterpart of neural network above, one can use a standard deep neural network where the primary objective is to generate the smooth $f_n$. In this framework, the layers situated between $z_n$ and $f_n$ are systematically constructed through a linear transformation from $z_n$ to $f_n$, leveraging the provided weights and activation functions: {see also footnote \ref{ftref}.} Such the deep neural network is denoted as ``Network 1" in Fig. \ref{SPANODE}. For further elaboration, we refer the readers to the comprehensive details in \cite{Hashimoto:2018ftp}.

\paragraph{More on the neural ODE approach.}
In order to further promote the algorithm of the discretized neural network \eqref{EM3} to its continuous counterpart \eqref{EM4}, we employ the \textit{PyTorch} framework: an open-source machine learning library. The ODE solver utilized for this purpose is \textit{torchdiffeq}, a submodule embedded within PyTorch designed specifically for neural ODE. In order to implement the neural ODE approach, one may also choose the adaptive ode solver. For instance, the ``DOPRI5" was employed for the holographic QCD within the neural ODE approach~\cite{Hashimoto:2020jug}. However, we employ a fixed step ODE solver rather than an adaptive one because it is faster and more practical.  See  Appendix \ref{AP2} for more details of our method and comparison between ODE solvers.

We proceed to elaborate on the methodology for training the bulk metric $f(z)$ using optical electric conductivity data. For numerical computations, we adopt the polynomial metric ansatz as introduced in \cite{Hashimoto:2020jug}:
\begin{align}\label{poly}
\begin{split}
    f(z) = \sum_{i=0}^4 b_i \, z^i \,,
\end{split}
\end{align}
where it involves five unknown coefficients to be trained, denoted as $(b_0,\,b_1,\,b_2,\,b_3,\,b_4)$. However, the imposition of the asymptotic AdS boundary condition $f(0)=1$ along with the black hole condition $f(1)=0$ leads to the constraints:
\begin{align}\label{coeff}
\begin{split}
    b_0 = 1\,, \qquad  b_4 = - (1 + b_1 + b_2 + b_3)  \,.
\end{split}
\end{align}
It is noteworthy that the ansatz \eqref{poly} in the bulk equations of motion \eqref{field_eq}, at the background level, results in the determination of $b_1=0$. This can be also observed in the analytic solution of the linear-axion model \eqref{true_metric}.\footnote{However, note that this may not hold true for the generalized holographic axion model~\cite{Baggioli:2021xuv}.} Essentially, with the neural ODE, we are left with training only two coefficients, namely $b_2$ and $b_3$.

\paragraph{Step 1.}
To initiate the algorithm, given $(\omega,\, \mu,\, \alpha)$, the random numbers $(A_{x}^{(1)},\,\Phi^{(1)})$ are set. Running the algorithm iteratively yields $(A_{x}^{(N)},\,\Phi^{(N)})$, and ultimately, $(A_{x}(z),\,\Phi(z))$ is obtained from the machine. The chosen number of layers in this paper is $N=10000$.

\paragraph{Step 2.}
The machine, having obtained $(A_{x}(z),\,\Phi(z))$, estimates the optical conductivity $\sigma_m(\omega)$ using the Kubo formula \eqref{conductivity} under the sourceless condition $\Phi(z_{fin})=0$ (or $\phi^{(S)}=0$ in \eqref{BCSET}). The estimation is given by:
\begin{align}
\begin{split}
    \sigma_m(\omega) = \frac{a_x'(z_{fin})}{i\omega a_x(z_{fin})} = \frac{A_x'(z_{fin})}{i\omega A_x(z_{fin})} - \frac{1}{f'(1)} \,.
\end{split}
\end{align}
The machine then can compare the estimated $\sigma_m(\omega)$ with the holographic conductivity $\sigma(\omega)$ at the same given $(\omega,\, \mu,\, \alpha)$.

\paragraph{Step 3.} 
The primary goal is to obtain the bulk metric function $f(z)$ or equivalently $(b_2,\,b_3)$ using optical conductivity data within the neural ODE method. For this purpose, the loss function $\mathcal{L}$ is defined as:
\begin{align}\label{loss}
\begin{split}
    \mathcal{L} = \frac{1}{N} \sum_{\omega} \Big[ | \sigma_m(\omega) - \sigma(\omega) | \,+\, \beta |\Phi(z_{fin})| \Big] \,.
\end{split}
\end{align}
Here, the first term in \eqref{loss} measures the difference between the output data $\sigma_m(\omega)$ and the true result $\sigma(\omega)$. The second term, the $\beta$ ``penalty" term, enforces the sourceless condition for utilizing the Kubo formula \eqref{conductivity}. The $\beta$ term plays a crucial role in shaping the loss function and influencing the optimization process.

Utilizing the \textit{Adam} optimization algorithm in the PyTorch machine learning framework, the optimized parameters $\left(A_{x}^{(1)},\,\Phi^{(1)}\right)$ can be found when the local minimum of $\mathcal{L}$ is attained by dialing $\beta$.\footnote{{For our numerical deep learning computations, we set the sourceless condition as $|\Phi(z_{fin})| \approx 10^{-4}$. For this purpose, we find $\beta=1$ in section \ref{sc3.2} and $\beta = 10^3$ in section \ref{sec33}.}} Subsequently, the trained bulk metric $f(z)$ is determined. In ideal situations, $\mathcal{L}$ is expected to be precisely zero, nevertheless in our numerical experiments, $\mathcal{L} \approx 10^{-2}$ is achieved.

In the subsequent subsection, employing the outlined algorithm and procedure, we determine the bulk metric $f(z)$ for two distinct scenarios. The first scenario involves utilizing boundary optical conductivity data derived from holography, while the second scenario involves utilizing data obtained from experiments of real material ($\text{UPd}_2\text{Al}_3$).

As an improved development of our methodology, we also extend our approach to ascertain the values of $(\mu,\, \alpha)$ from the machine, given the specified boundary conductivity data. In this context, the training parameters \(\theta\) encompass both $(\mu,\, \alpha)$ and $f(z)$.

%%%%%%%%%%%%
\subsection{Emergent spacetime from holographic conductivity}\label{sc3.2}
\begin{figure}[]
\centering
    \includegraphics[width=0.9\textwidth]{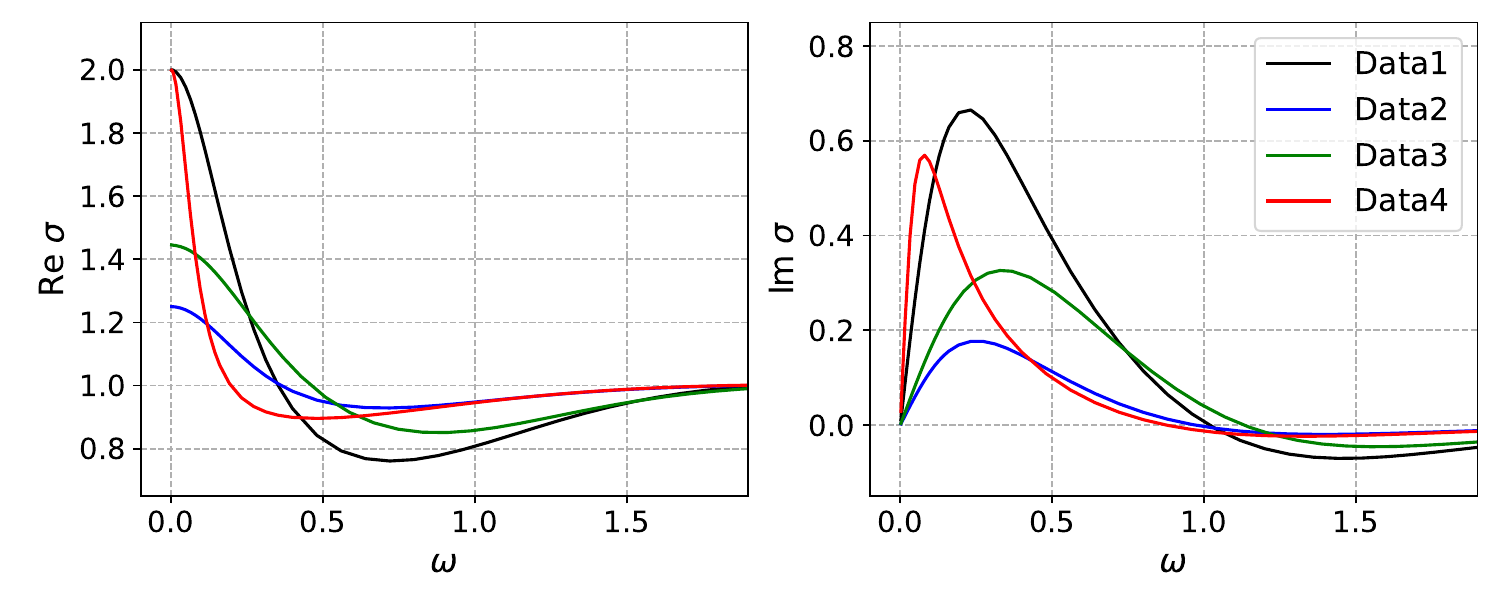}
     \caption{Optical electric conductivity of linear-axion model in holography. The data set is given in \eqref{data1_ref}. The left panel is the real part, while the right one is the imaginary part.}\label{fig:conductivity}
\end{figure}

Utilizing the aforementioned architecture, we conduct the training process. Specifically, we demonstrate that the spacetime can emerge from the ``holographic" electric conductivity data up to $\omega = 2$: See Fig. \ref{fig:conductivity} for the provided data set \eqref{data1_ref}.

\paragraph{Training $f(z)$ at given $(\mu,\,\alpha)$.} 
In this paper, we use the following four data as the specified parameters of interest
\begin{align}\label{data1_ref}
\begin{split}
\textrm{Data 1}:& \quad \mu = 1.0\,, \quad   \alpha = 1.0\,,  \qquad\,\,
\textrm{Data 2}: \quad \mu = 0.5\,, \quad   \alpha = 1.0\,,  \\
\textrm{Data 3}:& \quad \mu = 1.0\,, \quad   \alpha = 1.5\,, 
\qquad\,\,
\textrm{Data 4}: \quad \mu = 0.5\,, \quad   \alpha = 0.5\,.
\end{split}
\end{align}
Within our specified setup \eqref{data1_ref}, we have two training parameters ($b_2$ and $b_3$) for the metric function $f(z)$ in \eqref{poly} then.

\begin{figure}[]
\centering
    \includegraphics[width=0.5\textwidth]{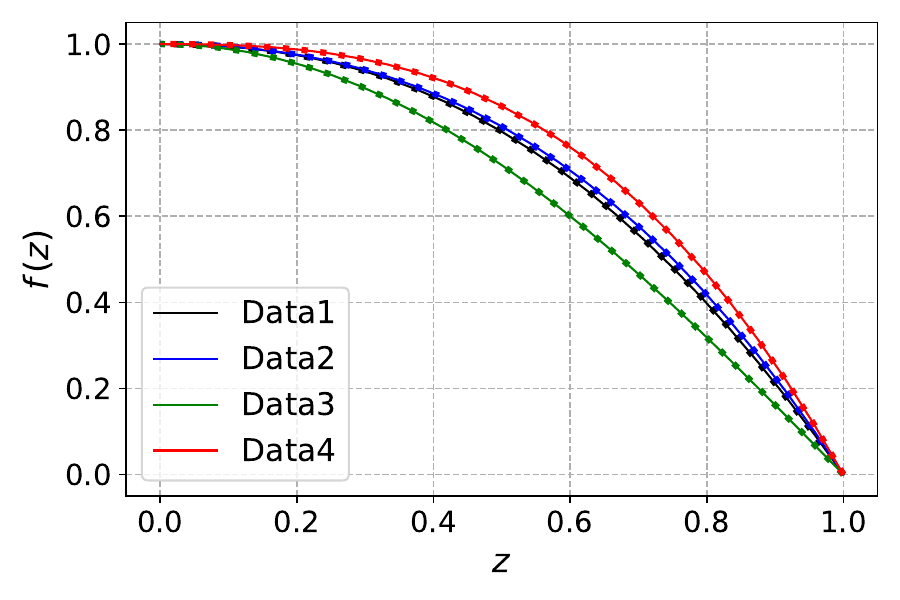}
     \caption{The metric function with the data set \eqref{data1_ref}. The dotted lines illustrate the metric corresponding to the linear-axion model \eqref{true_metric}, whereas the solid lines showcase the metric generated through machine learning.}\label{fig:metric}
\end{figure}

Our neural ODE approach successfully reveals the black hole geometry within the holographic bulk spacetime, as illustrated in Fig. \ref{fig:metric}. This emergent metric by the machine learning aligns with the geometry associated with the linear-axion model described by \eqref{true_metric}.
We also provide the obtained metric by the machine learning in Table. \ref{tab:compare}.
\begin{table}[]
    \centering
    \begin{tabular}{c||cl}
        %\hline
                &      & \qquad Metric function $f(z)$  \\\hline\hline
        Data 1  & \textit{True}         & $\:f(z;\mu,\alpha)\: = 1 - 0.5000 \, z^2 - 0.7500 \, z^3 + 0.2500 \, z^4$  \\
                & \textit{Trained} & $f(z;b_2,b_3) = 1 - 0.5048\, z^2 - 0.7435 \, z^3 + 0.2483 \, z^4$  \\\hline
        Data 2  & \textit{True}         & $\:f(z;\mu,\alpha)\: = 1 - 0.5000 \, z^2 - 0.5625 \, z^3 + 0.0625 \, z^4$  \\
                & \textit{Trained} & $f(z;b_2,b_3) = 1 - 0.5056\, z^2 - 0.5481 \, z^3 + 0.0537 \, z^4$  \\\hline
        Data 3  & \textit{True}         & $\:f(z;\mu,\alpha)\: = 1 - 1.1250 \,z^2 - 0.1250 \, z^3 + 0.2500 \, z^4$  \\
                & \textit{Trained} & $f(z;b_2,b_3) = 1 - 1.1294\, z^2 - 0.1146 \, z^3 + 0.2440 \, z^4$  \\\hline
        Data 4  & \textit{True}         & $\:f(z;\mu,\alpha)\: = 1 - 0.1250 \,z^2 - 0.9375 \,z^3 + 0.0625 \,z^4$  \\
                & \textit{Trained} & $f(z;b_2,b_3) = 1 - 0.1274\, z^2 - 0.9271 \, z^3 + 0.0544 \, z^4$  \\
    \end{tabular}
    \caption{The metric function with the data set \eqref{data1_ref}. Here, \textit{True} stands for the result of linear-axion model \eqref{true_metric}, while \textit{Trained} denotes the result by machine learning.}\label{tab:compare}
\end{table}
Our results suggest that the machine learning has the capability to precisely identify the metric based on the holographic conductivity data.

\paragraph{Training $f(z)$ and $(\mu,\,\alpha)$.} 
{To check the robustness and consistency of our method, we extend the training parameters: $(b_2, b_3) \rightarrow (b_2, b_3, \mu, \alpha)$.}
%Subsequently, we perform the neural ODE process to train the parameters $(\mu, \alpha)$ and $f(z)$. 
As an illustrative case, we employ machine learning using the optical conductivity data from Data 4 in Fig. \ref{fig:conductivity} without explicitly defining $(\mu, \alpha)$; for clarity, this case is denoted as Data 4*. In this scenario, the training variables encompass $(b_2, b_3, \mu, \alpha)$.

\begin{figure}[]
\centering
     \includegraphics[width=1\textwidth]{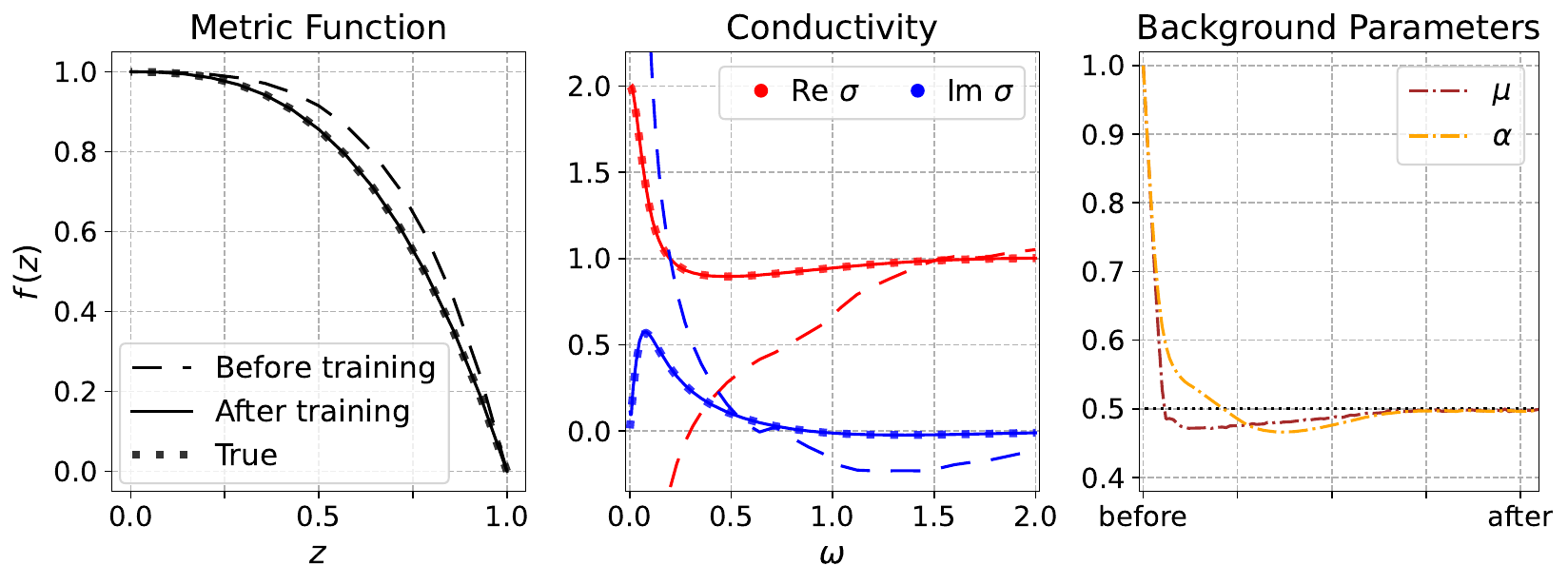}
     \caption{The result of Data 4* by the machine learning. The left figure displays the metric function, the center figure is the conductivity, and the right figure illustrates the parameters ($\mu, \alpha$). In each figure, the dashed line represents the machine learning data \textit{before} training, while the solid line corresponds to the data \textit{after} training. The dotted line in each case is the true data derived from the linear-axion model.}\label{fig:before_after}
\end{figure}

Our machine learning approach demonstrates not only successful reconstruction of the metric function of linear-axion model, but also the well trained parameters $(\mu, \alpha)$ as
\begin{align}
\begin{split}
(b_2,\,b_3) \,=\,
\begin{cases}
{(-0.1250,\,-0.9375)}     \quad    (\text{True})     \\
{(-0.1487,\,-0.8627)}     \quad    (\text{Trained})   \,, 
\end{cases}
\,\,
(\mu,\,\alpha) \,=\,
\begin{cases}
{(0.5000,\,0.5000)}       \quad    (\text{True})     \\
{(0.5004,\,0.4997)}       \quad    (\text{Trained})   \,,
\end{cases}
\,
\end{split}
\end{align}
which are closely align with Data 4 for $(\mu,\,\alpha)$ in \eqref{data1_ref}.
See also Fig. \ref{fig:before_after}.
It is evident that throughout the training procedure, both the metric function $f(z)$ and the accompanying parameters $(\mu, \alpha)$ progressively converge toward the values derived from the linear-axion model.

\begin{figure}[]
\centering
     \includegraphics[width=0.55\textwidth]{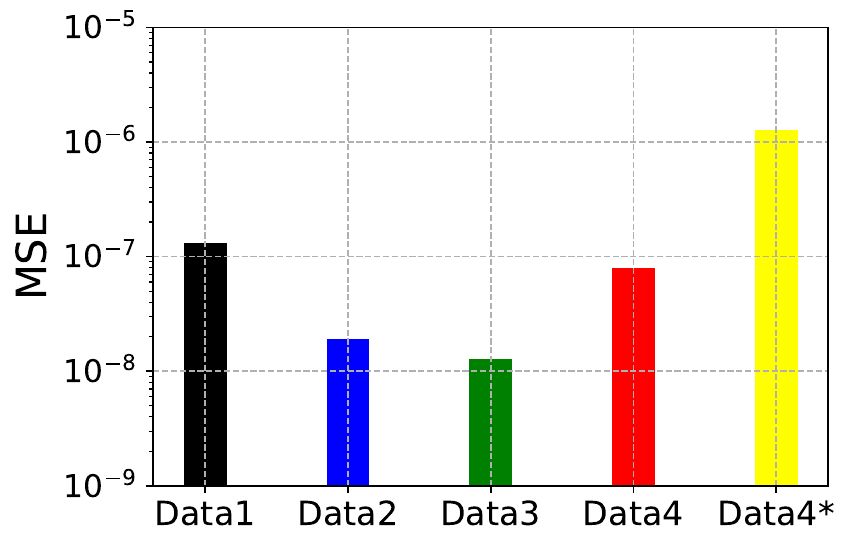}
     \caption{Mean Square Errors of the metric functions \eqref{MSEFOR} of all the data we have.}
     \label{fig:mse}
\end{figure}

Furthermore, we conduct an analysis of the Mean Square Error (MSE) between the metric functions of the linear-axion model and those obtained through machine learning, as illustrated in Fig. \ref{fig:mse}. 
The MSE is calculated as follows
\begin{equation}\label{MSEFOR}
    \textrm{MSE} = \frac{1}{N} \sum_z |f(z;\mu,\alpha) - f(z;b_2,b_3)|^2 \,.
\end{equation}
The MSE values are consistently below $10^{-7}$ for all data when the machine is provided with $(\mu,\,\alpha)$: i.e., from Data 1 to Data 4. Even without $(\mu,\,\alpha)$, Data 4*, the MSE remains on the order of $10^{-6}$.

%%%%%%%%%%%%
\subsection{Emergent spacetime from real material: $\text{UPd}_2\text{Al}_3$}\label{sec33}
In the preceding subsection, we demonstrated the applicability of deep learning in elucidating the AdS black hole spacetime within the context of the linear-axion model. Specifically, we illustrated that the spacetime can be emergent from the ``holographic" electric conductivity data.

Finally, we investigate a scenario involving the utilization of ``experimental" optical electric conductivity data obtained from real materials. For this purpose, we employ the data of $\text{UPd}_2\text{Al}_3$~\cite{Scheffler_2005}, which serves as a representative example of heavy fermion metals in strongly correlated electron systems.

\begin{figure}[]
\centering
    \includegraphics[width=0.8\textwidth]{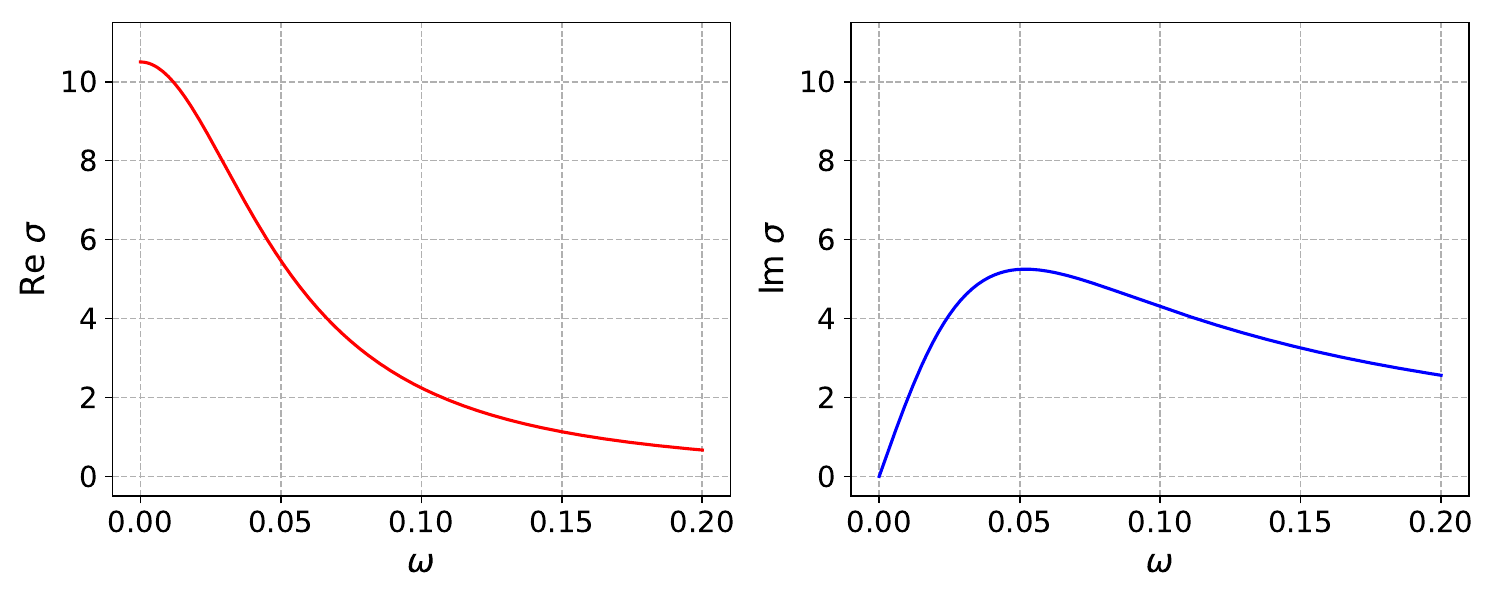}
     \caption{Optical conductivity of $\text{UPd}_2\text{Al}_3$ taken from \cite{Scheffler_2005}. The fitting is plotted by the Drude formula \eqref{DRMF} with the specified parameters \eqref{DRMF2}.}\label{DRUDEFIG}
\end{figure}

It is noteworthy that the expression governing the low-frequency conductivity in the Drude model is given by:
\begin{align}\label{DRMF}
\begin{split}
\sigma(\omega) = \frac{\sigma_0}{1 - i\omega \tau} \,,
\end{split}
\end{align}
where $\sigma_0$ is the DC conductivity, and $\tau$ denotes the relaxation time which is finite due to momentum relaxation.
In \cite{Scheffler_2005}, the authors established that the optical conductivity of the heavy fermion metal $\text{UPd}_2\text{Al}_3$ can be effectively described by the simple Drude model \eqref{DRMF} with the parameters
\begin{align}\label{DRMF2}
\begin{split}
\sigma_0 = 10.5 \,\mu\Omega^{-1} \text{m}^{-1} \,, \quad \tau = 4.8 \times 10^{-11} \,\text{s} \,.
\end{split}
\end{align}
We reproduce this conductivity plot in Fig. \ref{DRUDEFIG}, where $\tau$ and $\omega$ in \cite{Scheffler_2005} are rescaled by $10^{11}$ and $10^{-11}$ respectively so that they become numbers of order unity. This rescaling does not affect any physics but is useful for numerical analysis.\footnote{Specifically, for computational convenience in our deep learning process, we rescale $\omega$ as $4\omega \times 10^{11}$. Note that $\omega$ is scaled by $10^9$ in \cite{Scheffler_2005}.}

Before continuing, we address two remarks.
Firstly, there exists a general ambiguity regarding the relationship between physical quantities, such as ``$\omega$" utilized in holography, and their counterparts  $\omega$ in real word. This ambiguity exists in other parameters like temperature, chemical potential, and conductivity. Most holographic models, including our own, fall under the category of bottom-up models. In the absence of an exact top-down construction with a well-defined field theory dual, the interpretation of the ``frequency" (``$\omega$") remains ambiguous. 
%Moreover, even if these quantities are nominally identical, numerical differences may persist, such as $\omega \approx \# ``\omega"$. 
Therefore, it is advisable to establish a reference scale intrinsic to the model. For instance, in the context of superconductors, the phase transition temperature could be that case. Further insights into this discussion can be found in \cite{Jeong:2018tua,Ahn:2019lrh}. {Another way to fix this ambiguity is to determine the unknown bulk parameters such as the AdS radius, the Newton constant, bulk interaction(coupling) parameters or the location of horizon by the boundary experimental observables. 
In any case, for the purposes of our discussion, this ambiguity does not play any role, because it can be easily adjusted at the end by simple rescaling.}

%we assume that our holographic frequency ``$\omega$" aligns with the experimental $\omega_{\text{exp}}$.

Secondly, as demonstrated in \cite{Kim:2014bza}, the optical conductivity of the linear-axion model converges to a finite value in the high-frequency limit, which is in contrast with the Drude model \eqref{DRMF}. To avoid this issue, one may consider introducing additional contributions into \eqref{DRMF}, such as those arising from pair production~\cite{Kim:2014bza}. Such contributions can provide more accurate DC results as well. However, for our current focus on the low-frequency limit, where the original Drude formula \eqref{DRMF} is expected to perform well, we set aside these additional considerations. 

\begin{figure}[]
\centering
    \includegraphics[width=0.5\textwidth]{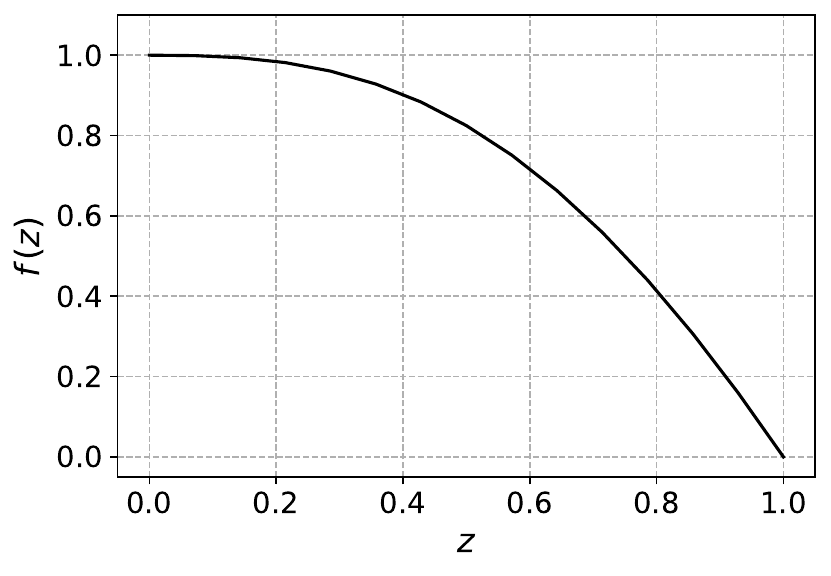}
     \caption{The emergent metric obtained by the machine learning from the data in Fig. \ref{DRUDEFIG}.}\label{fig:exp_metric}
\end{figure}

Employing the neural ODE method as previously detailed in the subsection, we now incorporate the optical electric conductivity from Fig. \ref{DRUDEFIG} as the boundary data. 
In particular, in order to assess the ability of the linear-axion model in characterizing the conductivity of the real material, we employ our polynomial ansatz \eqref{poly} to adhere the solution of the linear-axion model \eqref{true_metric}. In other words, the coefficients ($b_2, b_3$) are the functions of ($\mu,\,\alpha$) so we can train either ($b_2, b_3$) or ($\mu,\,\alpha$). After the deep-learning process,\footnote{{For computational convenience, we train variables ($\mu,\,\alpha$) instead of ($b_2, b_3$).
}} we obtain the emergent metric $f(z)$ shown in Fig. \ref{fig:exp_metric}, with the trained parameters
\begin{equation}\label{PRMSet}
   { b_2 = -0.1164 \,, \qquad b_3 = -1.4556 \,, }
\end{equation}
or equivalently
\begin{equation}\label{PRMSet}
   \mu = 1.5127\,, \qquad \alpha = 0.4825 \,.
\end{equation}
%
%Additionally, we also showcase $f(z)$ of the linear-axion model, \eqref{true_metric}, utilizing the parameters from \eqref{PRMSet}: 

\begin{figure}[]
\centering
     \includegraphics[width=0.9\textwidth]{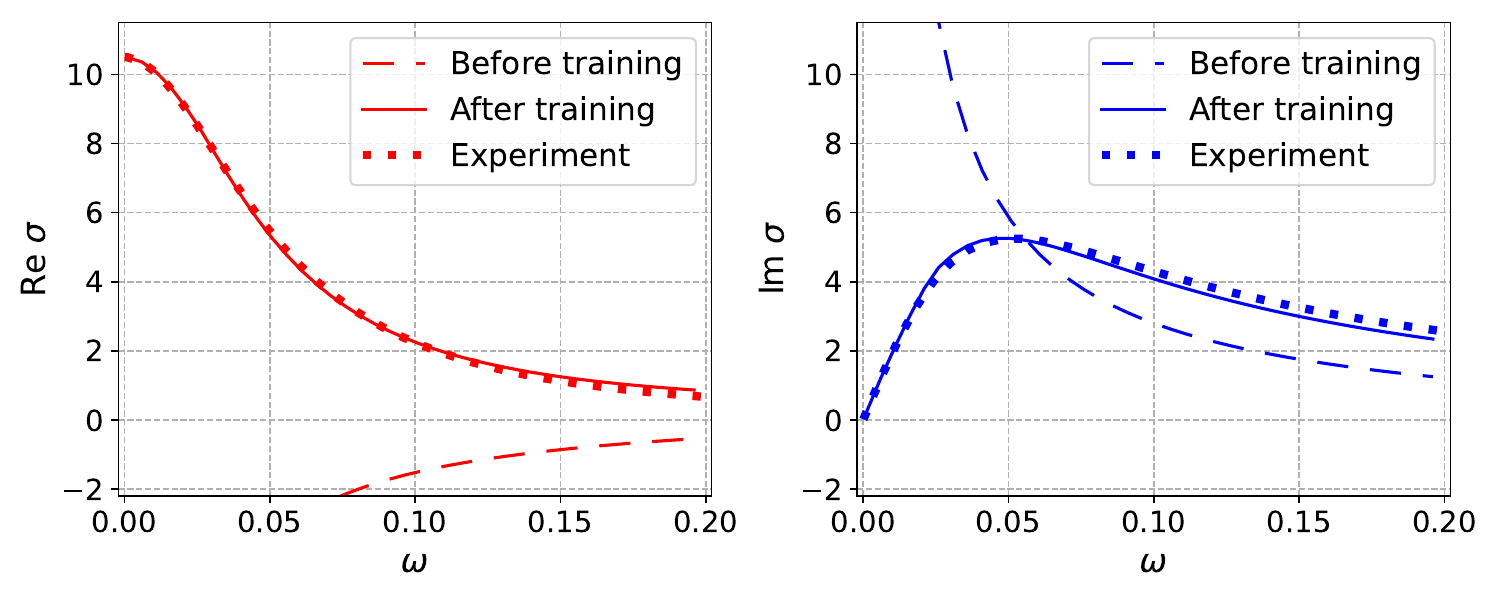}
     \caption{{The left (right) figure displays real (imaginary) part of the conductivity. In each figure, the dashed line represents the machine learning data \textit{before} training, while the solid line corresponds to the data \textit{after} training. The dotted line in each case is the experimental conductivity of $\text{UPd}_2\text{Al}_3$, i.e., Fig. \ref{DRUDEFIG}.}}\label{HOLOHOLO}
     %The left panel is the conductivity in the frequency regime up to $\omega=0.2$, while the right panel is the one for $\omega>0.2$.
\end{figure}

To check how precisely this trained metric yields the conductivity, we made comparison plots for conductivities in Fig.~\ref{HOLOHOLO}. The dashed lines are initial conductivities, which are very far from experimental data, dotted ones. However, after training, with the metric in Fig. \ref{fig:exp_metric}, the final conductivities (solid lines) are very close to the experimental data.
%\footnote{In the high-frequency regime, the real part of the optical conductivity of the linear-axion model \cite{Kim:2014bza} saturates to a finite value ($\text{Re} \, \sigma \approx 1$), while the imaginary part is vanishing ($\text{Im} \, \sigma \approx 0$)} 
To double-check, we also confirmed that the conductivities of the linear-axion model with \eqref{PRMSet} agree to the solid lines in Fig. \ref{HOLOHOLO}.
Our finding reveals that the deep learning with the boundary data in Fig. \ref{DRUDEFIG} may effectively discern the metric for the linear-axion model.

%\kyr{
%Given the inherent limitation of our machine learning computation, specifically tailored for solving equations within the linear-axion model, the machine tends to generate the emergent metric closely resembling that of the linear-axion model. Nevertheless, it can still remain uncertain whether such an emergent metric can authentically produce the conductivity of the real material.
%}

%To validate our machine learning results further, we also conducted an assessment of the ``holographic" optical electric conductivity using the obtained metric with \eqref{PRMSet}. The computed holographic conductivity is presented as dotted curves in Fig. \ref{HOLOHOLO}. They are very close to the experimental data, solid curves. 
%
%
%\kyr{
%The noticeable deviation, more pronounced beyond the low-frequency regime, is indicative of the insufficiency of the linear-axion model in precisely replicating the conductivity data of the real material.\footnote{We also find that a numerical instability in the $\omega=0$ limit results in a minor deviation in that region.} We further discuss this aspect in the concluding section.}
%
%Last but not least, we also make a plot of the conductivity across frequencies for $\omega>0.2$ in the right panel in Fig. \ref{HOLOHOLO} in order to illustrate the demonstration around \eqref{PRMSet}. 
%

%%%%%%%%%%%%
%%%%%%%%%%%%
\section{Conclusion}\label{sec4}
We have studied the bulk geometry of the AdS black hole within the framework of holographic condensed matter theory (CMT), with a focus on scenarios where translational symmetry is broken. Utilizing the neural ordinary differential equation (ODE)~\cite{Chen:2018aa} in the AdS/DL correspondence, we derived the bulk metric from the boundary electric conductivity data as a function of frequency. Our analysis pertains to a simple toy model, belonging to the typical class of holographic CMT models featuring broken translations: linear-axion models~\cite{Andrade:2013gsa, Baggioli:2021xuv}.
In essence, we examined a scenario with finite momentum relaxation to assess the effectiveness of machine-learning holographic CMT in a more realistic context.

The pioneering work of the neural ODE within the AdS/DL correspondence is explored in holographic QCD~\cite{Hashimoto:2020jug}, where the chiral condensate of lattice QCD serves as the boundary data.
It is notable that our study can signify the first application of the holographic CMT within the AdS/DL correspondence, employing the boundary optical conductivity data in the context of neural ODE. %Prior work involving neural ODE in holography is addressed in holographic QCD, as detailed in \cite{Hashimoto:2020jug}, where the chiral condensate of lattice QCD is employed as the boundary data.

It is worth noting that, before our paper, the authors in \cite{Li:2022zjc} deduced the bulk metric from conductivity using deep learning in holography. However, in that literature, the authors used so called reduced conductivity, which is not exclusively defined at the boundary but extends across the entire bulk dimension. 
In our research, there are three improvements compared to \cite{Li:2022zjc}. Firstly, we adopted the neural ODE methodology. By choosing a continuous function as the training variable, we not only ensured effective application to ODEs with complex structures but also enabled a more natural interpretation as the metric in spacetime. Secondly, we utilized optical conductivity data defined {\it strictly} at the boundary. By introducing regularity conditions into deep learning, we not only removed constraints in exploring data at various parameter values but also enabled a physically meaningful interpretation. Lastly, we considered the holographic model with the momentum relaxation, i.e., the translational symmetry is broken.

%However, there are three things to highlight for the distinctions. Firstly, their methodology does not involve neural ODE, implying that the corresponding metric is not continuous. Secondly, their optical conductivity, referred to as reduced conductivity, is not exclusively defined at the boundary but extends across the entire bulk dimension. Lastly, their holographic setup does not consider the momentum relaxation, i.e., the translational symmetry is not broken. 

%\bj{The authors in \cite{Li:2022zjc} also explored optical conductivity using deep learning in holography. It is worth noting that, in our research, there are three specific improvements compared to the previous study. Firstly, we adopted the Neural ODE methodology. By choosing a continuous function as the training variable, we not only ensured effective application to ODEs with complex structures but also enabled a more natural interpretation as the metric in spacetime. Secondly, we utilized optical conductivity data defined at the boundary in a commonly employed manner. By introducing regularity conditions into deep learning, we not only removed constraints in exploring data at various parameter values but also enabled a physically meaningful interpretation. Lastly, we considered the holographic model with the momentum relaxation, i.e., the translational symmetry is broken.}

In our work, we also improved the neural ODE framework used in previous studies on applied AdS/DL correspondence. Our modifications allow for the construction of neural networks based on the fluctuation bulk equations of motion for various fields (scalar, metric tensor, or gauge field), accounting for finite momentum relaxation and density. Importantly, our approach involves solving {\it coupled} bulk equations, which represents an enhancement over previous methods that focused on solving {\it single} bulk equations.

Following this, by stipulating that the ``holographic" optical electric conductivity serves as the boundary data for these coupled bulk equations of motion, we derived the metric for linear-axion models. This marks the successful completion of the bulk reconstruction program.

%We acknowledge a subtle issue and limitation in our methodology. While we utilized machine learning to reconstruct the metric from boundary conductivity data, the ultimate goal in holographic modeling of CMT is to identify \textit{the gravity action} corresponding to given CMT data, yielding the metric function as its solution. Determining the bulk action from its solution poses a nontrivial challenge, beyond the scope of this paper. Recent endeavors in holographic modeling of QCD, such as deriving the dilaton potential from chiral condensate~\cite{Hashimoto:2022eij}, highlight related efforts.

Additionally, we also explore a scenario involving the use of ``experimental" optical electric conductivity data from the real material $\text{UPd}_2\text{Al}_3$~\cite{Scheffler_2005}, representative of heavy fermion metals in strongly correlated electron systems. Employing the neural ODE method, we ascertain the bulk geometry of the linear-axion model through analysis of experimental data. However, our investigation holds validity within the small frequency regime, where the Drude model \eqref{DRMF} produces results consistent with the linear-axion model. 
%To clarify, beyond this small frequency range, the optical conductivity in the experiment data~\cite{Scheffler_2005} deviates from that of the linear-axion model. This suggests that our gravitational toy model (a linear-axion model) may be overly simplistic for capturing emergent spacetime from real materials.

In addition to employing more generalized gravitational models, such as the Q-lattice model, there are several avenues of extending our investigation to identify the proper gravitational dual corresponding to experimental data. 
Firstly, one can pursue the exploration involving a deep neural network\footnote{\label{ftref} A deep neural network is a sequence of continuous mappings and is described by $y^{(S)} = b^{(S-1)} + w^{(S-1)}\eta\left( b^{(S-2)} + w^{(S-2)}\eta\left( \cdots \eta\left( b^{(1)} + w^{(1)}y^{(1)}\right)\right)\right)$ where $S$ is the number of layers, $y^{(S)}$ is an output $f(z)$ corresponding to an input $y^{(1)}=z$, $w^{(k)}$ is a weight matrix for a linear mapping, $b^{(k)}$ is a bias vector for a translation, and $\eta$ is an activation function for a nonlinear mapping.
} 
without relying on the polynomial metric ansatz \eqref{poly}. The current polynomial ansatz may be overly restrictive, assuming continuity even before using the deep neural network. It is not guaranteed that the dual metric function aligns with such a polynomal form.
Secondly, beyond investigating the ``electric" optical conductivity, additional transport properties like thermal conductivity or thermoelectric conductivity can also be explored for the better training of the metric function. Additionally, the effect of the external magnetic field, e.g. Nernst effect, can also be beneficial in this context.
%
%As such, in this regard, our neural ODE approach may encounter challenges in producing an AdS black metric for the \textit{exactly} same data in \cite{Scheffler_2005}, i.e., for all the frequency range. This suggests that our gravitational toy model (a linear-axion model) may be overly simplistic for capturing emergent spacetime from real materials. We expect that successfully determining the gravity action from boundary data within the AdS/DL correspondence can guide the identification of an appropriate bulk action for real materials.

\paragraph{More on future research of holographic CMT within AdS/DL correspondence.}
Given the demonstrated feasibility of data-driven holographic modeling in CMT, the subsequent challenge is to identify a unified holographic CMT model capable of reproducing all the pertinent physical observable. To accomplish this, a comparison of various inversely-solved holographic models is essential, as briefly attempted with the toy model in this paper. Given the extensive amount of data in CMT, accomplishing unification may necessitate employing more sophisticated tools from the field of deep learning.

There can be several promising related directions worthy of future exploration. Undoubtedly, the primary and enduring motivation behind applying holography to condensed matter physics has been the pursuit of a deeper understanding of the enigmatic nature of the strange metal phase and its correlation with high-temperature superconductivity in materials such as cuprates and other strongly correlated systems~\cite{Liu_2012,Faulkner:2010aa}.

In this context, it is still plausible to employ optical conductivity, as demonstrated in this paper, to seek the corresponding holographic gravity model for cuprate materials. Specifically, one can consider leveraging well-established experimental universal features exhibited by such materials, such as the linear-in-temperature resistivity, Hall angle, and Homes' law in high-temperature superconductors.\footnote{For a recent discussion of the original AdS/CMT approach in this context, see \cite{Ahn:2023ciq}.} Also, one may endeavor to identify the gravity model for superconductors (in the context of AdS/DL) and compare it with the renowned holographic superconductors~\cite{Hartnoll:2008kx,Hartnoll:2008vx}.

Another intriguing avenue involving conductivity is the pursuit for the gravity dual of charge density waves~\cite{Gruner:1988zz}. This direction involves investigating scenarios where translational symmetry is pseudo-spontaneously broken. Such a setup gives rise to pinning phenomena, where impurities pin the phase Goldstone mode and produce a finite frequency peak in optical conductivity. This phenomenon is believed to play a significant role in the phase diagram of underdoped cuprate high-temperature superconductors.

Furthermore, beyond optical conductivity, studying the fermionic spectral function within the framework of AdS/DL is an interesting prospect. It is worth noting that the fermionic spectral function holds considerable importance, particularly in strongly coupled materials, and it can be directly probed experimentally through techniques such as Angle Resolved Photoemission Spectroscopy or Scanning Tunneling Microscopy. Several pioneering works~\cite{Cubrovic:2009ye,Faulkner:2010zz,Lee:2008xf,Liu:2009dm,Faulkner:2009wj} have explored the fermionic spectral function in the context of holography, shedding light on potential non-Fermi liquid signatures.

We leave these topics for future investigation and plan to address them in the near future.

%-----------------------

%\paragraph{AdS/Neural ODE and Condensed Matter Theory}
%\sout{The AdS/DL framework has been applied to condensed matter theory by \cite{Li:2022zjc}. They introduced a vector field model for optical conductivity and solved ODE using \ref{EM3}. The study confirmed the effectiveness of AdS/DL, even when the field depends on another variable $\omega$ in addition to $z$. As a toy model, the researchers generated horizon data from the conductivity using the true model and utilized it as a solution data.

%Neural ODE have been introduced to AdS/DL by \cite{Hashimoto:2020jug}. They solved the ODE of the scalar field $\phi(z)$ to determine the metric from lattice QCD data.}

%Thanks to the work of \cite{Li:2022zjc,Hashimoto:2020jug}, \sout{we adopt the neural ODE presented in \eqref{EM4} instead of \eqref{EM3} and verify that the neural ODE also performs well, even when the field depends on another variable $\omega$ in addition to $z$. Furthermore, we extend the model to the linear-axion model with the horizon conditions \ref{FOCOEFF} for bulk fields $A_x$ and $\Phi$. As a result, we must solve coupled ODEs, and we confirm that AdS/DL and neural ODE still work well for coupled ODEs.}

%%%%%%%%%%%%%%%%%%%%%%%%%%%%%%%%
%    Section: Acknowledgments
%%%%%%%%%%%%%%%%%%%%%%%%%%%%%%%%
\acknowledgments
%We would like to thank {} for valuable discussions and correspondence.
%
This work was supported by the Basic Science Research Program through the National Research Foundation of Korea (NRF) funded by the Ministry of Science, ICT $\&$ Future Planning (NRF-2021R1A2C1006791) and the AI-based GIST Research Scientist Project grant funded by the GIST in 2023. This work was also supported by Creation
of the Quantum Information Science R$\&$D Ecosystem (Grant No. 2022M3H3A106307411)
through the National Research Foundation of Korea (NRF) funded by the Korean government (Ministry of Science and ICT).
H.-S Jeong acknowledges the support of the Spanish MINECO ``Centro de Excelencia Severo Ochoa'' Programme under grant SEV-2012-0249. This work is supported through the grants CEX2020-001007-S and PID2021-123017NB-I00, funded by MCIN/AEI/10.13039/501100011033 and by ERDF A way of making Europe.
B. Ahn was supported by Basic Science Research Program through the National Research Foundation of Korea funded by the Ministry of Education (NRF-2020R1A6A3A01095962, NRF-2022R1I1A1A01064342).
All the authors contributed equally to this paper and should be considered as co-first authors.

%%%%%%%%%%%%%%%%%%%%%%%%%%%%%%%%%%%%%%%%%%%%%%%%%%%
%  Appendix
%%%%%%%%%%%%%%%%%%%%%%%%%%%%%%%%%%%%%%%%%%%%%%%%%%%

\appendix
\section{4th Order Runge Kutta Method}\label{AP1}
Runge-Kutta method is a numerical method for solving ordinary differential equations. This method is developed by Carl Runge and Wilhelm Kutta. Most simple method to solve an initial value problem is using Euler method.
\begin{equation}
    y_{n+1} = y_n + \Delta t \cdot h(t_n, y_n) \,.
\end{equation}
Euler method has a single slope in each step. However, 4th order Runge-Kutta method has four different slopes in each step as the following:
\begin{equation} \label{rk4}
    y_{n+1} = y_n + \frac{\Delta t}{6} (k_1 + 2k_2 + 2k_3 + k_4) \,,
\end{equation}
where
\begin{align}
\begin{split}
    &k_1 = h(t_n, y_n)\,,\\
    &k_2 = h(t_n + \frac{\Delta t}{2}, \; y_n + \frac{\Delta t}{2} \cdot k_1) \,,\\
    &k_3 = h(t_n + \frac{\Delta t}{2}, \; y_n + \frac{\Delta t}{2} \cdot k_2) \,,\\
    &k_4 = h(t_n + \Delta t, \; y_n + \Delta t \cdot k_3)\,.
\end{split}
\end{align}

For example, we consider $y_n = [A'_n, A_n]$ and $t_n = z_n$ such that $A''_n$ is given by the linear combination of $A'_n$ and $A_n$. Then, the slope $h(t,y):= \partial_t \, y$ is given by
\begin{equation}
    h(t_n,y_n) = \Big[ P(t_n)A'_n + Q(t_n)A_n, \;\; A'_n \Big] \,,
\end{equation}
where $P(t_n)$ and $Q(t_n)$ are the any coefficient functions to make $A''_n$. That is a reason $y_{n+1}$ gets the linear combination of $A'_n$ and $A_n$ by \eqref{rk4}.
\begin{equation}
    y_{n+1} = \Big[ \Tilde{P}(t_n,\Delta t)A'_n + \Tilde{Q}(t_n,\Delta t)A_n, \;\; \Tilde{R}(t_n,\Delta t)A'_n + \Tilde{S}(t_n,\Delta t)A_n \Big] \,.
\end{equation}
Note that the neural network consists of the linear combination with bias $b$ and activation function $\eta$.
\begin{equation}
    y_{n+1} = \eta\big(W_n y_n + b_n\big) \,.
\end{equation}
If we fix $b = 0$ and specify $\eta$ as identity function, then the weight is
\begin{equation}
    W_n = \begin{pmatrix}
        \Tilde{P}(t_n,\Delta t) && \Tilde{Q}(t_n,\Delta t) \\
        \Tilde{R}(t_n,\Delta t) && \Tilde{S}(t_n,\Delta t)
        \end{pmatrix} \,.
\end{equation}
Therefore, the final value $y_N$ is given by 
\begin{equation}\label{RK4_linear}
    y_N = \prod_{n=0}^{N-1} W_n \cdot y_0 \,.
\end{equation}

\section{Adaptive ODE Solvers}\label{AP2}
ODE solvers are numerical methods used to find solutions to differential equations. The numerical methods include Euler's method, Runge-Kutta method, and various adaptive solvers. The adaptive solvers adjust the size of steps automatically.\footnote{Here, the step means the discretized coordinate on the propagation direction like $z$ in Fig. \ref{SPANODE}.} For adaptive step size, adaptive solvers typically employ two different numerical methods. In almost adaptive solvers, two distinct numerical methods are a higher order Runge-Kutta method and a lower order Runge-Kutta method.\footnote{For example, DOPRI5 method, most preferred ODE solver, employs 4th order and 5th order Runge-Kutta method.} The solver calculates one step using both methods and estimates an error $\epsilon$ between the results of both. 
\begin{equation}
    \epsilon^{(n+1)} = | y_1^{(n+1)} - y_2^{(n+1)}| \;, \quad y_k^{(n+1)} = y^{(n)} + R_k(z^{(n)}, y^{(n)})\cdot\Delta z \;\;\; (k=1,2) \,,
\end{equation}
where the upper index $n$ means the step and lower one separates two Runge-Kutta methods. $R_k$ is given by the order of Runge-Kutta method, like \eqref{rk4}.
If the error is larger than a predefined tolerance, the solver reduces the step size $\Delta z$. In contrary, if the error is smaller than the tolerance, the solver increases it. By adopting this adaptive method, the ODE solvers guarantee a certain level of accuracy, but they may come with significant computing costs and time.

The adaptive solvers require significant time in our problem. Hence, we employ the fixed step ODE solver instead of adaptive one. The step size $\Delta z$ is specified in each step through the uniform distribution of $\tilde{z} = \log(1-z+\delta)$ to make the step size smaller near the horizon. We introduce a shift factor $\delta=0.1$ to ensure a minimum number of steps near the AdS boundary. 

\begin{table}[]
    \centering
    \begin{tabular}{c|c|r|r|c}
         ODE Solver &  Cutoff  & 10 epochs   & 100 epochs    &  $b_2$  \\\hline\hline
         Our solver &   $10^{-4}$   &       3 min &        37 min & -0.5015 \\\hline
         DOPRI5     &   $10^{-4}$   & 17 hr\;\;9 min &       -       &    -    \\
                    &   $10^{-3}$   &      14 min & 2 hr\;\;\;6 min & -0.5157 \\
                    &   $10^{-2}$   &       1 min &        16 min & -0.5847 \\\hline
         Fehlberg2  &   $10^{-4}$   &      23 min & 3 hr   38 min & -0.5024 \\
                    &   $10^{-3}$   &       2 min &        20 min & -0.5140 \\
                    &   $10^{-2}$   &       1 min &        14 min & -0.5856 \\
    \end{tabular}
    \caption{Comparison of learning results of Data\,1 over time according to ODE solver. Epoch refers to the number of times the training parameters have been updated. Cutoff means UV and IR cutoffs, as shown in \eqref{cutoff}.}
    \label{comp_solver}
\end{table}

For comparison with adaptive solvers, we simplify the machine learning problem compared to section \ref{sc3.2}. The Hawking temperature can be obtained by \eqref{tem}. Note that the Hawking temperature, derived from \eqref{poly} to \eqref{coeff}, can also be expressed as 
\begin{equation}\label{a3}
    T_H = - \frac{f'(1)}{4\pi} = \frac{4+2b_2+b_3}{4\pi} \,.
\end{equation}
In other words, within our specified setup \eqref{data1_ref}, the metric has a single training parameter denoted as $b_2$, with $b_3$ determined based on the $T_H$. In this case, Data\,1 is learned using the three ODE solvers, separately. One is our fixed step solver with 2000 steps, while the others are adaptive ODE solvers as known as DOPRI5 and Fehlberg2.\footnote{In section \ref{sc3.2}, the number of steps is 10000. However, the other solvers also require more computing costs as the number of training parameters increases. In this sense, our solver remains more efficient than the others.} As shown in Table. \ref{comp_solver}, our solver demonstrated the highest accuracy with affordable computing costs.

%%%%%%%%%%%%%%%%%%%%%%%%%%%%%%%%%%%%%%%%%%%%%%%%%%%
%  END
%%%%%%%%%%%%%%%%%%%%%%%%%%%%%%%%%%%%%%%%%%%%%%%%%%%

%\bibliography{Refs}

\begin{thebibliography}{100}

\bibitem{Ammon:2015wua}
M.~Ammon and J.~Erdmenger, \emph{{Gauge/gravity duality}}.
\newblock Cambridge Univ. Pr., Cambridge, UK, 2015.

\bibitem{Hartnoll:2016apf}
S.~A. Hartnoll, A.~Lucas and S.~Sachdev, \emph{{Holographic quantum matter}},
  \href{http://arxiv.org/abs/1612.07324}{{\tt 1612.07324}}.

\bibitem{Hartnoll:2009sz}
S.~A. Hartnoll, \emph{{Lectures on holographic methods for condensed matter
  physics}},
  \href{http://dx.doi.org/10.1088/0264-9381/26/22/224002}{\emph{Class.Quant.Grav.}
  {\bf 26} (2009) 224002}, [\href{http://arxiv.org/abs/0903.3246}{{\tt
  0903.3246}}].

\bibitem{Zaanen:2015oix}
J.~Zaanen, Y.-W. Sun, Y.~Liu and K.~Schalm, \emph{{Holographic Duality in
  Condensed Matter Physics}}.
\newblock Cambridge Univ. Press, 2015.

\bibitem{CasalderreySolana:2011us}
J.~Casalderrey-Solana, H.~Liu, D.~Mateos, K.~Rajagopal and U.~A. Wiedemann,
  \emph{{Gauge/String Duality, Hot QCD and Heavy Ion Collisions}},
  \href{http://arxiv.org/abs/1101.0618}{{\tt 1101.0618}}.

\bibitem{Baggioli:2019rrs}
M.~Baggioli, \emph{{Applied Holography}: {A Practical Mini-Course}},  other
  thesis, Madrid, IFT, 2019.
\newblock 10.1007/978-3-030-35184-7.

\bibitem{Heller:2016gbp}
M.~P. Heller, \emph{{Holography, Hydrodynamization and Heavy-Ion Collisions}},
  \href{http://dx.doi.org/10.5506/APhysPolB.47.2581}{\emph{Acta Phys. Polon. B}
  {\bf 47} (2016) 2581}, [\href{http://arxiv.org/abs/1610.02023}{{\tt
  1610.02023}}].

\bibitem{Florkowski:2017olj}
W.~Florkowski, M.~P. Heller and M.~Spalinski, \emph{{New theories of
  relativistic hydrodynamics in the LHC era}},
  \href{http://dx.doi.org/10.1088/1361-6633/aaa091}{\emph{Rept. Prog. Phys.}
  {\bf 81} (2018) 046001}, [\href{http://arxiv.org/abs/1707.02282}{{\tt
  1707.02282}}].

\bibitem{HBT2020}
M.~Schirber, \emph{{Holographist by Trade}}, {\emph{The American Physical
  Society} (2020) }.

\bibitem{Chen:2021lnq}
B.~Chen, B.~Czech and Z.-z. Wang, \emph{{Quantum information in holographic
  duality}}, \href{http://dx.doi.org/10.1088/1361-6633/ac51b5}{\emph{Rept.
  Prog. Phys.} {\bf 85} (2022) 046001},
  [\href{http://arxiv.org/abs/2108.09188}{{\tt 2108.09188}}].

\bibitem{Maldacena:1997re}
J.~M. Maldacena, \emph{{The Large N limit of superconformal field theories and
  supergravity}}, \href{http://dx.doi.org/10.1023/A:1026654312961,
  10.1023/A:1026654312961}{\emph{Adv.Theor.Math.Phys.} {\bf 2} (1998)
  231--252}, [\href{http://arxiv.org/abs/hep-th/9711200}{{\tt
  hep-th/9711200}}].

\bibitem{Aharony:1999ti}
O.~Aharony, S.~S. Gubser, J.~M. Maldacena, H.~Ooguri and Y.~Oz, \emph{{Large N
  field theories, string theory and gravity}},
  \href{http://dx.doi.org/10.1016/S0370-1573(99)00083-6}{\emph{Phys. Rept.}
  {\bf 323} (2000) 183--386}, [\href{http://arxiv.org/abs/hep-th/9905111}{{\tt
  hep-th/9905111}}].

\bibitem{Gruner:1988zz}
G.~Gruner, \emph{{The dynamics of charge-density waves}},
  \href{http://dx.doi.org/10.1103/RevModPhys.60.1129}{\emph{Rev. Mod. Phys.}
  {\bf 60} (1988) 1129--1181}.

\bibitem{Donos:2012js}
A.~Donos and S.~A. Hartnoll, \emph{{Interaction-driven localization in
  holography}}, \href{http://dx.doi.org/10.1038/nphys2701}{\emph{Nature Phys.}
  {\bf 9} (2013) 649--655}, [\href{http://arxiv.org/abs/1212.2998}{{\tt
  1212.2998}}].

\bibitem{Donos:2014oha}
A.~Donos, B.~Gout{\'e}raux and E.~Kiritsis, \emph{{Holographic Metals and
  Insulators with Helical Symmetry}},
  \href{http://arxiv.org/abs/1406.6351}{{\tt 1406.6351}}.

\bibitem{Donos:2014gya}
A.~Donos, J.~P. Gauntlett and C.~Pantelidou, \emph{{Conformal field theories in
  $d=4$ with a helical twist}},  \href{http://arxiv.org/abs/1412.3446}{{\tt
  1412.3446}}.

\bibitem{Donos:2013eha}
A.~Donos and J.~P. Gauntlett, \emph{{Holographic Q-lattices}},
  \href{http://dx.doi.org/10.1007/JHEP04(2014)040}{\emph{JHEP} {\bf 1404}
  (2014) 040}, [\href{http://arxiv.org/abs/1311.3292}{{\tt 1311.3292}}].

\bibitem{Donos:2014uba}
A.~Donos and J.~P. Gauntlett, \emph{{Novel metals and insulators from
  holography}}, \href{http://dx.doi.org/10.1007/JHEP06(2014)007}{\emph{JHEP}
  {\bf 1406} (2014) 007}, [\href{http://arxiv.org/abs/1401.5077}{{\tt
  1401.5077}}].

\bibitem{Andrade:2013gsa}
T.~Andrade and B.~Withers, \emph{{A simple holographic model of momentum
  relaxation}}, \href{http://dx.doi.org/10.1007/JHEP05(2014)101}{\emph{JHEP}
  {\bf 1405} (2014) 101}, [\href{http://arxiv.org/abs/1311.5157}{{\tt
  1311.5157}}].

\bibitem{Baggioli:2021xuv}
M.~Baggioli, K.-Y. Kim, L.~Li and W.-J. Li, \emph{{Holographic Axion Model: a
  simple gravitational tool for quantum matter}},
  \href{http://dx.doi.org/10.1007/s11433-021-1681-8}{\emph{Sci. China Phys.
  Mech. Astron.} {\bf 64} (2021) 270001},
  [\href{http://arxiv.org/abs/2101.01892}{{\tt 2101.01892}}].

\bibitem{Vegh:2013sk}
D.~Vegh, \emph{{Holography without translational symmetry}},
  \href{http://arxiv.org/abs/1301.0537}{{\tt 1301.0537}}.

\bibitem{Davison:2013jba}
R.~A. Davison, \emph{{Momentum relaxation in holographic massive gravity}},
  \href{http://dx.doi.org/10.1103/PhysRevD.88.086003}{\emph{Phys.Rev.} {\bf
  D88} (2013) 086003}, [\href{http://arxiv.org/abs/1306.5792}{{\tt
  1306.5792}}].

\bibitem{Blake:2013bqa}
M.~Blake and D.~Tong, \emph{{Universal Resistivity from Holographic Massive
  Gravity}},
  \href{http://dx.doi.org/10.1103/PhysRevD.88.106004}{\emph{Phys.Rev.} {\bf
  D88} (2013) 106004}, [\href{http://arxiv.org/abs/1308.4970}{{\tt
  1308.4970}}].

\bibitem{Blake:2013owa}
M.~Blake, D.~Tong and D.~Vegh, \emph{{Holographic Lattices Give the Graviton a
  Mass}},
  \href{http://dx.doi.org/10.1103/PhysRevLett.112.071602}{\emph{Phys.Rev.Lett.}
  {\bf 112} (2014) 071602}, [\href{http://arxiv.org/abs/1310.3832}{{\tt
  1310.3832}}].

\bibitem{Alberte:2015isw}
L.~Alberte, M.~Baggioli, A.~Khmelnitsky and O.~Pujolas, \emph{{Solid Holography
  and Massive Gravity}},
  \href{http://dx.doi.org/10.1007/JHEP02(2016)114}{\emph{JHEP} {\bf 02} (2016)
  114}, [\href{http://arxiv.org/abs/1510.09089}{{\tt 1510.09089}}].

\bibitem{Horowitz:2012ky}
G.~T. Horowitz, J.~E. Santos and D.~Tong, \emph{{Optical Conductivity with
  Holographic Lattices}},
  \href{http://dx.doi.org/10.1007/JHEP07(2012)168}{\emph{JHEP} {\bf 1207}
  (2012) 168}, [\href{http://arxiv.org/abs/1204.0519}{{\tt 1204.0519}}].

\bibitem{Horowitz:2012gs}
G.~T. Horowitz, J.~E. Santos and D.~Tong, \emph{{Further Evidence for
  Lattice-Induced Scaling}},
  \href{http://dx.doi.org/10.1007/JHEP11(2012)102}{\emph{JHEP} {\bf 1211}
  (2012) 102}, [\href{http://arxiv.org/abs/1209.1098}{{\tt 1209.1098}}].

\bibitem{Nakamura:2009tf}
S.~Nakamura, H.~Ooguri and C.-S. Park, \emph{{Gravity Dual of Spatially
  Modulated Phase}},
  \href{http://dx.doi.org/10.1103/PhysRevD.81.044018}{\emph{Phys. Rev.} {\bf
  D81} (2010) 044018}, [\href{http://arxiv.org/abs/0911.0679}{{\tt
  0911.0679}}].

\bibitem{Donos:2011bh}
A.~Donos and J.~P. Gauntlett, \emph{{Holographic striped phases}},
  \href{http://dx.doi.org/10.1007/JHEP08(2011)140}{\emph{JHEP} {\bf 08} (2011)
  140}, [\href{http://arxiv.org/abs/1106.2004}{{\tt 1106.2004}}].

\bibitem{Ahn:2022azl}
Y.~Ahn, M.~Baggioli, K.-B. Huh, H.-S. Jeong, K.-Y. Kim and Y.-W. Sun,
  \emph{{Holography and magnetohydrodynamics with dynamical gauge fields}},
  \href{http://dx.doi.org/10.1007/JHEP02(2023)012}{\emph{JHEP} {\bf 02} (2023)
  012}, [\href{http://arxiv.org/abs/2211.01760}{{\tt 2211.01760}}].

\bibitem{Jeong:2023las}
H.-S. Jeong, M.~Baggioli, K.-Y. Kim and Y.-W. Sun, \emph{{Collective dynamics
  and the Anderson-Higgs mechanism in a bona fide holographic superconductor}},
  \href{http://dx.doi.org/10.1007/JHEP03(2023)206}{\emph{JHEP} {\bf 03} (2023)
  206}, [\href{http://arxiv.org/abs/2302.02364}{{\tt 2302.02364}}].

\bibitem{Davison:2013txa}
R.~A. Davison, K.~Schalm and J.~Zaanen, \emph{{Holographic duality and the
  resistivity of strange metals}},
  \href{http://dx.doi.org/10.1103/PhysRevB.89.245116}{\emph{Phys. Rev.} {\bf
  B89} (2014) 245116}, [\href{http://arxiv.org/abs/1311.2451}{{\tt
  1311.2451}}].

\bibitem{Gouteraux:2014hca}
B.~Gout{\'e}raux, \emph{{Charge transport in holography with momentum
  dissipation}}, \href{http://dx.doi.org/10.1007/JHEP04(2014)181}{\emph{JHEP}
  {\bf 1404} (2014) 181}, [\href{http://arxiv.org/abs/1401.5436}{{\tt
  1401.5436}}].

\bibitem{Blauvelt:2017koq}
E.~Blauvelt, S.~Cremonini, A.~Hoover, L.~Li and S.~Waskie, \emph{{Holographic
  model for the anomalous scalings of the cuprates}},
  \href{http://dx.doi.org/10.1103/PhysRevD.97.061901}{\emph{Phys. Rev.} {\bf
  D97} (2018) 061901}, [\href{http://arxiv.org/abs/1710.01326}{{\tt
  1710.01326}}].

\bibitem{Alberte:2017cch}
L.~Alberte, M.~Ammon, M.~Baggioli, A.~Jimnez and O.~Pujol~s, \emph{{Black hole
  elasticity and gapped transverse phonons in holography}},
  \href{http://dx.doi.org/10.1007/JHEP01(2018)129}{\emph{JHEP} {\bf 01} (2018)
  129}, [\href{http://arxiv.org/abs/1708.08477}{{\tt 1708.08477}}].

\bibitem{Jeong:2018tua}
H.-S. Jeong, K.-Y. Kim and C.~Niu, \emph{{Linear-$T$ resistivity at high
  temperature}}, \href{http://dx.doi.org/10.1007/JHEP10(2018)191}{\emph{JHEP}
  {\bf 10} (2018) 191}, [\href{http://arxiv.org/abs/1806.07739}{{\tt
  1806.07739}}].

\bibitem{PhysRevLett.120.171602}
L.~Alberte, M.~Ammon, A.~Jim\'enez-Alba, M.~Baggioli and O.~Pujol\`as,
  \emph{Holographic phonons},
  \href{http://dx.doi.org/10.1103/PhysRevLett.120.171602}{\emph{Phys. Rev.
  Lett.} {\bf 120} (Apr, 2018) 171602}.

\bibitem{Ammon:2019wci}
M.~Ammon, M.~Baggioli and A.~Jimenez-Alba, \emph{{A Unified Description of
  Translational Symmetry Breaking in Holography}},
  \href{http://arxiv.org/abs/1904.05785}{{\tt 1904.05785}}.

\bibitem{Ahn:2019lrh}
Y.~Ahn, H.-S. Jeong, D.~Ahn and K.-Y. Kim, \emph{{Linear-$T$ resistivity from
  low to high temperature: axion-dilaton theories}},
  \href{http://dx.doi.org/10.1007/JHEP04(2020)153}{\emph{JHEP} {\bf 04} (2020)
  153}, [\href{http://arxiv.org/abs/1907.12168}{{\tt 1907.12168}}].

\bibitem{Jeong:2021wiu}
H.-S. Jeong and K.-Y. Kim, \emph{{Homes\textquoteright{} law in holographic
  superconductor with linear-T resistivity}},
  \href{http://dx.doi.org/10.1007/JHEP03(2022)060}{\emph{JHEP} {\bf 03} (2022)
  060}, [\href{http://arxiv.org/abs/2112.01153}{{\tt 2112.01153}}].

\bibitem{Baggioli:2022pyb}
M.~Baggioli and B.~Gout\'eraux, \emph{{Colloquium: Hydrodynamics and holography
  of charge density wave phases}},
  \href{http://dx.doi.org/10.1103/RevModPhys.95.011001}{\emph{Rev. Mod. Phys.}
  {\bf 95} (2023) 011001}, [\href{http://arxiv.org/abs/2203.03298}{{\tt
  2203.03298}}].

\bibitem{Ahn:2023ciq}
Y.~Ahn, M.~Baggioli, H.-S. Jeong and K.-Y. Kim, \emph{{Inability of linear
  axion holographic Gubser-Rocha model to capture all the transport anomalies
  of strange metals}},
  \href{http://dx.doi.org/10.1103/PhysRevB.108.235104}{\emph{Phys. Rev. B} {\bf
  108} (2023) 235104}, [\href{http://arxiv.org/abs/2307.04433}{{\tt
  2307.04433}}].

\bibitem{Jeong:2019zab}
H.-S. Jeong, K.-Y. Kim, Y.~Seo, S.-J. Sin and S.-Y. Wu, \emph{{Holographic
  Spectral Functions with Momentum Relaxation}},
  \href{http://dx.doi.org/10.1103/PhysRevD.102.026017}{\emph{Phys. Rev. D} {\bf
  102} (2020) 026017}, [\href{http://arxiv.org/abs/1910.11034}{{\tt
  1910.11034}}].

\bibitem{Davison:2014lua}
R.~A. Davison and B.~Gout\'eraux, \emph{{Momentum dissipation and effective
  theories of coherent and incoherent transport}},
  \href{http://dx.doi.org/10.1007/JHEP01(2015)039}{\emph{JHEP} {\bf 01} (2015)
  039}, [\href{http://arxiv.org/abs/1411.1062}{{\tt 1411.1062}}].

\bibitem{Blake:2016jnn}
M.~Blake and A.~Donos, \emph{{Diffusion and Chaos from near AdS$_2$ horizons}},
  \href{http://dx.doi.org/10.1007/JHEP02(2017)013}{\emph{JHEP} {\bf 02} (2017)
  013}, [\href{http://arxiv.org/abs/1611.09380}{{\tt 1611.09380}}].

\bibitem{Blake:2017qgd}
M.~Blake, R.~A. Davison and S.~Sachdev, \emph{{Thermal diffusivity and chaos in
  metals without quasiparticles}},
  \href{http://dx.doi.org/10.1103/PhysRevD.96.106008}{\emph{Phys. Rev. D} {\bf
  96} (2017) 106008}, [\href{http://arxiv.org/abs/1705.07896}{{\tt
  1705.07896}}].

\bibitem{Baggioli:2017ojd}
M.~Baggioli and W.-J. Li, \emph{{Diffusivities bounds and chaos in holographic
  Horndeski theories}},
  \href{http://dx.doi.org/10.1007/JHEP07(2017)055}{\emph{JHEP} {\bf 07} (2017)
  055}, [\href{http://arxiv.org/abs/1705.01766}{{\tt 1705.01766}}].

\bibitem{Ahn:2017kvc}
H.-S. Jeong, Y.~Ahn, D.~Ahn, C.~Niu, W.-J. Li and K.-Y. Kim, \emph{{Thermal
  diffusivity and butterfly velocity in anisotropic Q-Lattice models}},
  \href{http://dx.doi.org/10.1007/JHEP01(2018)140}{\emph{JHEP} {\bf 01} (2018)
  140}, [\href{http://arxiv.org/abs/1708.08822}{{\tt 1708.08822}}].

\bibitem{Davison:2018ofp}
R.~A. Davison, S.~A. Gentle and B.~Gout\'eraux, \emph{{Slow relaxation and
  diffusion in holographic quantum critical phases}},
  \href{http://dx.doi.org/10.1103/PhysRevLett.123.141601}{\emph{Phys. Rev.
  Lett.} {\bf 123} (2019) 141601}, [\href{http://arxiv.org/abs/1808.05659}{{\tt
  1808.05659}}].

\bibitem{Blake:2018leo}
M.~Blake, R.~A. Davison, S.~Grozdanov and H.~Liu, \emph{{Many-body chaos and
  energy dynamics in holography}},
  \href{http://dx.doi.org/10.1007/JHEP10(2018)035}{\emph{JHEP} {\bf 10} (2018)
  035}, [\href{http://arxiv.org/abs/1809.01169}{{\tt 1809.01169}}].

\bibitem{Arean:2020eus}
D.~Arean, R.~A. Davison, B.~Gout\'eraux and K.~Suzuki, \emph{{Hydrodynamic
  Diffusion and Its Breakdown near AdS2 Quantum Critical Points}},
  \href{http://dx.doi.org/10.1103/PhysRevX.11.031024}{\emph{Phys. Rev. X} {\bf
  11} (2021) 031024}, [\href{http://arxiv.org/abs/2011.12301}{{\tt
  2011.12301}}].

\bibitem{Liu:2021qmt}
Y.~Liu and X.-M. Wu, \emph{{Breakdown of hydrodynamics from holographic pole
  collision}}, \href{http://dx.doi.org/10.1007/JHEP01(2022)155}{\emph{JHEP}
  {\bf 01} (2022) 155}, [\href{http://arxiv.org/abs/2111.07770}{{\tt
  2111.07770}}].

\bibitem{Jeong:2021zhz}
H.-S. Jeong, K.-Y. Kim and Y.-W. Sun, \emph{{Bound of diffusion constants from
  pole-skipping points: spontaneous symmetry breaking and magnetic field}},
  \href{http://dx.doi.org/10.1007/JHEP07(2021)105}{\emph{JHEP} {\bf 07} (2021)
  105}, [\href{http://arxiv.org/abs/2104.13084}{{\tt 2104.13084}}].

\bibitem{Wu:2021mkk}
N.~Wu, M.~Baggioli and W.-J. Li, \emph{{On the universality of AdS$_{2}$
  diffusion bounds and the breakdown of linearized hydrodynamics}},
  \href{http://dx.doi.org/10.1007/JHEP05(2021)014}{\emph{JHEP} {\bf 05} (2021)
  014}, [\href{http://arxiv.org/abs/2102.05810}{{\tt 2102.05810}}].

\bibitem{Jeong:2021zsv}
H.-S. Jeong, K.-Y. Kim and Y.-W. Sun, \emph{{The breakdown of
  magneto-hydrodynamics near AdS$_{2}$ fixed point and energy diffusion
  bound}}, \href{http://dx.doi.org/10.1007/JHEP02(2022)006}{\emph{JHEP} {\bf
  02} (2022) 006}, [\href{http://arxiv.org/abs/2105.03882}{{\tt 2105.03882}}].

\bibitem{Huh:2021ppg}
K.-B. Huh, H.-S. Jeong, K.-Y. Kim and Y.-W. Sun, \emph{{Upper bound of the
  charge diffusion constant in holography}},
  \href{http://dx.doi.org/10.1007/JHEP07(2022)013}{\emph{JHEP} {\bf 07} (2022)
  013}, [\href{http://arxiv.org/abs/2111.07515}{{\tt 2111.07515}}].

\bibitem{Baggioli:2022uqb}
M.~Baggioli, S.~Grieninger, S.~Grozdanov and Z.~Lu, \emph{{Aspects of
  univalence in holographic axion models}},
  \href{http://dx.doi.org/10.1007/JHEP11(2022)032}{\emph{JHEP} {\bf 11} (2022)
  032}, [\href{http://arxiv.org/abs/2205.06076}{{\tt 2205.06076}}].

\bibitem{Jeong:2023ynk}
H.-S. Jeong, \emph{{Quantum Chaos and Pole-Skipping in Semi-Locally Critical
  IR}},  \href{http://arxiv.org/abs/2309.13412}{{\tt 2309.13412}}.

\bibitem{Grozdanov:2017ajz}
S.~Grozdanov, K.~Schalm and V.~Scopelliti, \emph{{Black hole scrambling from
  hydrodynamics}},
  \href{http://dx.doi.org/10.1103/PhysRevLett.120.231601}{\emph{Phys. Rev.
  Lett.} {\bf 120} (2018) 231601}, [\href{http://arxiv.org/abs/1710.00921}{{\tt
  1710.00921}}].

\bibitem{Blake:2017ris}
M.~Blake, H.~Lee and H.~Liu, \emph{{A quantum hydrodynamical description for
  scrambling and many-body chaos}},
  \href{http://dx.doi.org/10.1007/JHEP10(2018)127}{\emph{JHEP} {\bf 10} (2018)
  127}, [\href{http://arxiv.org/abs/1801.00010}{{\tt 1801.00010}}].

\bibitem{Grozdanov:2019uhi}
S.~Grozdanov, P.~K. Kovtun, A.~O. Starinets and P.~Tadi\'c, \emph{{The complex
  life of hydrodynamic modes}},
  \href{http://dx.doi.org/10.1007/JHEP11(2019)097}{\emph{JHEP} {\bf 11} (2019)
  097}, [\href{http://arxiv.org/abs/1904.12862}{{\tt 1904.12862}}].

\bibitem{Blake:2019otz}
M.~Blake, R.~A. Davison and D.~Vegh, \emph{{Horizon constraints on holographic
  Green\textquoteright{}s functions}},
  \href{http://dx.doi.org/10.1007/JHEP01(2020)077}{\emph{JHEP} {\bf 01} (2020)
  077}, [\href{http://arxiv.org/abs/1904.12883}{{\tt 1904.12883}}].

\bibitem{Natsuume:2019xcy}
M.~Natsuume and T.~Okamura, \emph{{Nonuniqueness of Green\textquoteright{}s
  functions at special points}},
  \href{http://dx.doi.org/10.1007/JHEP12(2019)139}{\emph{JHEP} {\bf 12} (2019)
  139}, [\href{http://arxiv.org/abs/1905.12015}{{\tt 1905.12015}}].

\bibitem{Natsuume:2019sfp}
M.~Natsuume and T.~Okamura, \emph{{Holographic chaos, pole-skipping, and
  regularity}}, \href{http://dx.doi.org/10.1093/ptep/ptz155}{\emph{PTEP} {\bf
  2020} (2020) 013B07}, [\href{http://arxiv.org/abs/1905.12014}{{\tt
  1905.12014}}].

\bibitem{Natsuume:2019vcv}
M.~Natsuume and T.~Okamura, \emph{{Pole-skipping with finite-coupling
  corrections}},
  \href{http://dx.doi.org/10.1103/PhysRevD.100.126012}{\emph{Phys. Rev. D} {\bf
  100} (2019) 126012}, [\href{http://arxiv.org/abs/1909.09168}{{\tt
  1909.09168}}].

\bibitem{Ceplak:2019ymw}
N.~Ceplak, K.~Ramdial and D.~Vegh, \emph{{Fermionic pole-skipping in
  holography}}, \href{http://dx.doi.org/10.1007/JHEP07(2020)203}{\emph{JHEP}
  {\bf 07} (2020) 203}, [\href{http://arxiv.org/abs/1910.02975}{{\tt
  1910.02975}}].

\bibitem{Ahn:2019rnq}
Y.~Ahn, V.~Jahnke, H.-S. Jeong and K.-Y. Kim, \emph{{Scrambling in Hyperbolic
  Black Holes: shock waves and pole-skipping}},
  \href{http://dx.doi.org/10.1007/JHEP10(2019)257}{\emph{JHEP} {\bf 10} (2019)
  257}, [\href{http://arxiv.org/abs/1907.08030}{{\tt 1907.08030}}].

\bibitem{Ahn:2020bks}
Y.~Ahn, V.~Jahnke, H.-S. Jeong, K.-Y. Kim, K.-S. Lee and M.~Nishida,
  \emph{{Pole-skipping of scalar and vector fields in hyperbolic space:
  conformal blocks and holography}},
  \href{http://dx.doi.org/10.1007/JHEP09(2020)111}{\emph{JHEP} {\bf 09} (2020)
  111}, [\href{http://arxiv.org/abs/2006.00974}{{\tt 2006.00974}}].

\bibitem{Abbasi:2020ykq}
N.~Abbasi and S.~Tahery, \emph{{Complexified quasinormal modes and the
  pole-skipping in a holographic system at finite chemical potential}},
  \href{http://dx.doi.org/10.1007/JHEP10(2020)076}{\emph{JHEP} {\bf 10} (2020)
  076}, [\href{http://arxiv.org/abs/2007.10024}{{\tt 2007.10024}}].

\bibitem{Liu:2020yaf}
Y.~Liu and A.~Raju, \emph{{Quantum Chaos in Topologically Massive Gravity}},
  \href{http://dx.doi.org/10.1007/JHEP12(2020)027}{\emph{JHEP} {\bf 12} (2020)
  027}, [\href{http://arxiv.org/abs/2005.08508}{{\tt 2005.08508}}].

\bibitem{Ramirez:2020qer}
D.~M. Ramirez, \emph{{Chaos and pole skipping in CFT$_{2}$}},
  \href{http://dx.doi.org/10.1007/JHEP12(2021)006}{\emph{JHEP} {\bf 12} (2021)
  006}, [\href{http://arxiv.org/abs/2009.00500}{{\tt 2009.00500}}].

\bibitem{Ahn:2020baf}
Y.~Ahn, V.~Jahnke, H.-S. Jeong, K.-Y. Kim, K.-S. Lee and M.~Nishida,
  \emph{{Classifying pole-skipping points}},
  \href{http://dx.doi.org/10.1007/JHEP03(2021)175}{\emph{JHEP} {\bf 03} (2021)
  175}, [\href{http://arxiv.org/abs/2010.16166}{{\tt 2010.16166}}].

\bibitem{Natsuume:2020snz}
M.~Natsuume and T.~Okamura, \emph{{Pole-skipping and zero temperature}},
  \href{http://dx.doi.org/10.1103/PhysRevD.103.066017}{\emph{Phys. Rev. D} {\bf
  103} (2021) 066017}, [\href{http://arxiv.org/abs/2011.10093}{{\tt
  2011.10093}}].

\bibitem{Ceplak:2021efc}
N.~Ceplak and D.~Vegh, \emph{{Pole-skipping and Rarita-Schwinger fields}},
  \href{http://dx.doi.org/10.1103/PhysRevD.103.106009}{\emph{Phys. Rev. D} {\bf
  103} (2021) 106009}, [\href{http://arxiv.org/abs/2101.01490}{{\tt
  2101.01490}}].

\bibitem{Natsuume:2021fhn}
M.~Natsuume and T.~Okamura, \emph{{Nonuniqueness of scattering amplitudes at
  special points}},
  \href{http://dx.doi.org/10.1103/PhysRevD.104.126007}{\emph{Phys. Rev. D} {\bf
  104} (2021) 126007}, [\href{http://arxiv.org/abs/2108.07832}{{\tt
  2108.07832}}].

\bibitem{Blake:2021hjj}
M.~Blake and R.~A. Davison, \emph{{Chaos and pole-skipping in rotating black
  holes}}, \href{http://dx.doi.org/10.1007/JHEP01(2022)013}{\emph{JHEP} {\bf
  01} (2022) 013}, [\href{http://arxiv.org/abs/2111.11093}{{\tt 2111.11093}}].

\bibitem{Jeong:2022luo}
H.-S. Jeong, K.-Y. Kim and Y.-W. Sun, \emph{{Quasi-normal modes of dyonic black
  holes and magneto-hydrodynamics}},
  \href{http://dx.doi.org/10.1007/JHEP07(2022)065}{\emph{JHEP} {\bf 07} (2022)
  065}, [\href{http://arxiv.org/abs/2203.02642}{{\tt 2203.02642}}].

\bibitem{Wang:2022mcq}
D.~Wang and Z.-Y. Wang, \emph{{Pole Skipping in Holographic Theories with
  Bosonic Fields}},
  \href{http://dx.doi.org/10.1103/PhysRevLett.129.231603}{\emph{Phys. Rev.
  Lett.} {\bf 129} (2022) 231603}, [\href{http://arxiv.org/abs/2208.01047}{{\tt
  2208.01047}}].

\bibitem{Amano:2022mlu}
M.~A.~G. Amano, M.~Blake, C.~Cartwright, M.~Kaminski and A.~P. Thompson,
  \emph{{Chaos and pole-skipping in a simply spinning plasma}},
  \href{http://dx.doi.org/10.1007/JHEP02(2023)253}{\emph{JHEP} {\bf 02} (2023)
  253}, [\href{http://arxiv.org/abs/2211.00016}{{\tt 2211.00016}}].

\bibitem{Yuan:2023tft}
H.~Yuan, X.-H. Ge, K.-Y. Kim, C.-W. Ji and Y.~Ahn, \emph{{Pole-skipping points
  in 2D gravity and SYK model}},
  \href{http://dx.doi.org/10.1007/JHEP08(2023)157}{\emph{JHEP} {\bf 08} (2023)
  157}, [\href{http://arxiv.org/abs/2303.04801}{{\tt 2303.04801}}].

\bibitem{Grozdanov:2023txs}
S.~Grozdanov and M.~Vrbica, \emph{{Pole-skipping of gravitational waves in the
  backgrounds of four-dimensional massive black holes}},
  \href{http://arxiv.org/abs/2303.15921}{{\tt 2303.15921}}.

\bibitem{Natsuume:2023lzy}
M.~Natsuume and T.~Okamura, \emph{{Pole skipping in a non-black-hole
  geometry}}, \href{http://dx.doi.org/10.1103/PhysRevD.108.046012}{\emph{Phys.
  Rev. D} {\bf 108} (2023) 046012},
  [\href{http://arxiv.org/abs/2306.03930}{{\tt 2306.03930}}].

\bibitem{Ning:2023pmz}
S.~Ning, D.~Wang and Z.-Y. Wang, \emph{{Pole skipping in holographic theories
  with gauge and fermionic fields}},
  \href{http://arxiv.org/abs/2308.08191}{{\tt 2308.08191}}.

\bibitem{Grozdanov:2023tag}
S.~Grozdanov, T.~Lemut and J.~F. Pedraza, \emph{{Reconstruction of the
  quasinormal spectrum from pole-skipping}},
  \href{http://arxiv.org/abs/2308.01371}{{\tt 2308.01371}}.

\bibitem{Jeong:2023rck}
H.-S. Jeong, C.-W. Ji and K.-Y. Kim, \emph{{Pole-skipping in rotating BTZ black
  holes}}, \href{http://dx.doi.org/10.1007/JHEP08(2023)139}{\emph{JHEP} {\bf
  08} (2023) 139}, [\href{http://arxiv.org/abs/2306.14805}{{\tt 2306.14805}}].

\bibitem{Natsuume:2023hsz}
M.~Natsuume and T.~Okamura, \emph{{Pole skipping as missing states}},
  \href{http://dx.doi.org/10.1103/PhysRevD.108.106006}{\emph{Phys. Rev. D} {\bf
  108} (2023) 106006}, [\href{http://arxiv.org/abs/2307.11178}{{\tt
  2307.11178}}].

\bibitem{Abbasi:2023myj}
N.~Abbasi and K.~Landsteiner, \emph{{Pole-skipping as order parameter to probe
  a quantum critical point}},
  \href{http://dx.doi.org/10.1007/JHEP09(2023)169}{\emph{JHEP} {\bf 09} (2023)
  169}, [\href{http://arxiv.org/abs/2307.16716}{{\tt 2307.16716}}].

\bibitem{RezaMohammadiMozaffar:2016lbo}
M.~Reza Mohammadi~Mozaffar, A.~Mollabashi and F.~Omidi, \emph{{Non-local Probes
  in Holographic Theories with Momentum Relaxation}},
  \href{http://dx.doi.org/10.1007/JHEP10(2016)135}{\emph{JHEP} {\bf 10} (2016)
  135}, [\href{http://arxiv.org/abs/1608.08781}{{\tt 1608.08781}}].

\bibitem{Yekta:2020wup}
D.~M. Yekta, H.~Babaei-Aghbolagh, K.~Babaei~Velni and H.~Mohammadzadeh,
  \emph{{Holographic complexity for black branes with momentum relaxation}},
  \href{http://dx.doi.org/10.1103/PhysRevD.104.086025}{\emph{Phys. Rev. D} {\bf
  104} (2021) 086025}, [\href{http://arxiv.org/abs/2009.01340}{{\tt
  2009.01340}}].

\bibitem{Li:2019rpp}
Y.-Z. Li and X.-M. Kuang, \emph{{Probes of holographic thermalization in a
  simple model with momentum relaxation}},
  \href{http://dx.doi.org/10.1016/j.nuclphysb.2020.115043}{\emph{Nucl. Phys. B}
  {\bf 956} (2020) 115043}, [\href{http://arxiv.org/abs/1911.11980}{{\tt
  1911.11980}}].

\bibitem{Zhou:2019xzc}
Y.-T. Zhou, X.-M. Kuang, Y.-Z. Li and J.-P. Wu, \emph{{Holographic subregion
  complexity under a thermal quench in an Einstein-Maxwell-axion theory with
  momentum relaxation}},
  \href{http://dx.doi.org/10.1103/PhysRevD.101.106024}{\emph{Phys. Rev. D} {\bf
  101} (2020) 106024}, [\href{http://arxiv.org/abs/1912.03479}{{\tt
  1912.03479}}].

\bibitem{Huang:2019zph}
Y.-f. Huang, Z.-j. Shi, C.~Niu, C.-y. Zhang and P.~Liu, \emph{{Mixed State
  Entanglement for Holographic Axion Model}},
  \href{http://dx.doi.org/10.1140/epjc/s10052-020-7921-y}{\emph{Eur. Phys. J.
  C} {\bf 80} (2020) 426}, [\href{http://arxiv.org/abs/1911.10977}{{\tt
  1911.10977}}].

\bibitem{Jeong:2022zea}
H.-S. Jeong, K.-Y. Kim and Y.-W. Sun, \emph{{Holographic entanglement density
  for spontaneous symmetry breaking}},
  \href{http://dx.doi.org/10.1007/JHEP06(2022)078}{\emph{JHEP} {\bf 06} (2022)
  078}, [\href{http://arxiv.org/abs/2203.07612}{{\tt 2203.07612}}].

\bibitem{HosseiniMansoori:2022hok}
S.~A. Hosseini~Mansoori, O.~Luongo, S.~Mancini, M.~Mirjalali, M.~Rafiee and
  A.~Tavanfar, \emph{{Planar black holes in holographic axion gravity: Islands,
  Page times, and scrambling times}},
  \href{http://dx.doi.org/10.1103/PhysRevD.106.126018}{\emph{Phys. Rev. D} {\bf
  106} (2022) 126018}, [\href{http://arxiv.org/abs/2209.00253}{{\tt
  2209.00253}}].

\bibitem{Chen:2018aa}
R.~T.~Q. Chen, Y.~Rubanova, J.~Bettencourt and D.~Duvenaud, \emph{Neural
  ordinary differential equations},
  \href{http://arxiv.org/abs/1806.07366}{{\tt 1806.07366}}.

\bibitem{Hashimoto:2018ftp}
K.~Hashimoto, S.~Sugishita, A.~Tanaka and A.~Tomiya, \emph{{Deep learning and
  the AdS/CFT correspondence}},
  \href{http://dx.doi.org/10.1103/PhysRevD.98.046019}{\emph{Phys. Rev. D} {\bf
  98} (2018) 046019}, [\href{http://arxiv.org/abs/1802.08313}{{\tt
  1802.08313}}].

\bibitem{You:2017guh}
Y.-Z. You, Z.~Yang and X.-L. Qi, \emph{{Machine Learning Spatial Geometry from
  Entanglement Features}},
  \href{http://dx.doi.org/10.1103/PhysRevB.97.045153}{\emph{Phys. Rev. B} {\bf
  97} (2018) 045153}, [\href{http://arxiv.org/abs/1709.01223}{{\tt
  1709.01223}}].

\bibitem{Hashimoto:2018bnb}
K.~Hashimoto, S.~Sugishita, A.~Tanaka and A.~Tomiya, \emph{{Deep Learning and
  Holographic QCD}},
  \href{http://dx.doi.org/10.1103/PhysRevD.98.106014}{\emph{Phys. Rev. D} {\bf
  98} (2018) 106014}, [\href{http://arxiv.org/abs/1809.10536}{{\tt
  1809.10536}}].

\bibitem{Hu:2019nea}
H.-Y. Hu, S.-H. Li, L.~Wang and Y.-Z. You, \emph{{Machine Learning Holographic
  Mapping by Neural Network Renormalization Group}},
  \href{http://dx.doi.org/10.1103/PhysRevResearch.2.023369}{\emph{Phys. Rev.
  Res.} {\bf 2} (2020) 023369}, [\href{http://arxiv.org/abs/1903.00804}{{\tt
  1903.00804}}].

\bibitem{Hashimoto:2019bih}
K.~Hashimoto, \emph{{AdS/CFT correspondence as a deep Boltzmann machine}},
  \href{http://dx.doi.org/10.1103/PhysRevD.99.106017}{\emph{Phys. Rev. D} {\bf
  99} (2019) 106017}, [\href{http://arxiv.org/abs/1903.04951}{{\tt
  1903.04951}}].

\bibitem{Han:2019wue}
X.~Han and S.~A. Hartnoll, \emph{{Deep Quantum Geometry of Matrices}},
  \href{http://dx.doi.org/10.1103/PhysRevX.10.011069}{\emph{Phys. Rev. X} {\bf
  10} (2020) 011069}, [\href{http://arxiv.org/abs/1906.08781}{{\tt
  1906.08781}}].

\bibitem{Tan:2019czc}
J.~Tan and C.-B. Chen, \emph{{Deep learning the holographic black hole with
  charge}}, \href{http://dx.doi.org/10.1142/S0218271819501530}{\emph{Int. J.
  Mod. Phys. D} {\bf 28} (2019) 1950153},
  [\href{http://arxiv.org/abs/1908.01470}{{\tt 1908.01470}}].

\bibitem{Yan:2020wcd}
Y.-K. Yan, S.-F. Wu, X.-H. Ge and Y.~Tian, \emph{{Deep learning black hole
  metrics from shear viscosity}},
  \href{http://dx.doi.org/10.1103/PhysRevD.102.101902}{\emph{Phys. Rev. D} {\bf
  102} (4, 2020) 101902}, [\href{http://arxiv.org/abs/2004.12112}{{\tt
  2004.12112}}].

\bibitem{Akutagawa:2020yeo}
T.~Akutagawa, K.~Hashimoto and T.~Sumimoto, \emph{{Deep Learning and AdS/QCD}},
  \href{http://dx.doi.org/10.1103/PhysRevD.102.026020}{\emph{Phys. Rev. D} {\bf
  102} (2020) 026020}, [\href{http://arxiv.org/abs/2005.02636}{{\tt
  2005.02636}}].

\bibitem{Hashimoto:2020jug}
K.~Hashimoto, H.-Y. Hu and Y.-Z. You, \emph{{Neural ordinary differential
  equation and holographic quantum chromodynamics}},
  \href{http://dx.doi.org/10.1088/2632-2153/abe527}{\emph{Mach. Learn. Sci.
  Tech.} {\bf 2} (2021) 035011}, [\href{http://arxiv.org/abs/2006.00712}{{\tt
  2006.00712}}].

\bibitem{Chen:2020dxg}
H.-Y. Chen, Y.-H. He, S.~Lal and M.~Z. Zaz, \emph{{Machine Learning Etudes in
  Conformal Field Theories}},  \href{http://arxiv.org/abs/2006.16114}{{\tt
  2006.16114}}.

\bibitem{Song:2020agw}
M.~Song, M.~S.~H. Oh, Y.~Ahn and K.-Y. Kim, \emph{{AdS/Deep-Learning made easy:
  simple examples}},
  \href{http://dx.doi.org/10.1088/1674-1137/abfc36}{\emph{Chin. Phys. C} {\bf
  45} (2021) 073111}, [\href{http://arxiv.org/abs/2011.13726}{{\tt
  2011.13726}}].

\bibitem{Hashimoto:2021ihd}
K.~Hashimoto, K.~Ohashi and T.~Sumimoto, \emph{{Deriving the dilaton potential
  in improved holographic QCD from the meson spectrum}},
  \href{http://dx.doi.org/10.1103/PhysRevD.105.106008}{\emph{Phys. Rev. D} {\bf
  105} (2022) 106008}, [\href{http://arxiv.org/abs/2108.08091}{{\tt
  2108.08091}}].

\bibitem{Lam:2021ugb}
J.~Lam and Y.-Z. You, \emph{{Machine learning statistical gravity from
  multi-region entanglement entropy}},
  \href{http://dx.doi.org/10.1103/PhysRevResearch.3.043199}{\emph{Phys. Rev.
  Res.} {\bf 3} (2021) 043199}, [\href{http://arxiv.org/abs/2110.01115}{{\tt
  2110.01115}}].

\bibitem{Park:2022fqy}
C.~Park, C.-O. Hwang, K.~Cho and S.-J. Kim, \emph{{Dual geometry of
  entanglement entropy via deep learning}},
  \href{http://dx.doi.org/10.1103/PhysRevD.106.106017}{\emph{Phys. Rev. D} {\bf
  106} (2022) 106017}, [\href{http://arxiv.org/abs/2205.04445}{{\tt
  2205.04445}}].

\bibitem{Katsube:2022ofz}
R.~Katsube, W.-H. Tam, M.~Hotta and Y.~Nambu, \emph{{Deep learning metric
  detectors in general relativity}},
  \href{http://dx.doi.org/10.1103/PhysRevD.106.044051}{\emph{Phys. Rev. D} {\bf
  106} (2022) 044051}, [\href{http://arxiv.org/abs/2206.03006}{{\tt
  2206.03006}}].

\bibitem{Hashimoto:2022eij}
K.~Hashimoto, K.~Ohashi and T.~Sumimoto, \emph{{Deriving the dilaton potential
  in improved holographic QCD from the chiral condensate}},
  \href{http://dx.doi.org/10.1093/ptep/ptad026}{\emph{PTEP} {\bf 2023} (2023)
  033B01}, [\href{http://arxiv.org/abs/2209.04638}{{\tt 2209.04638}}].

\bibitem{Li:2022zjc}
K.~Li, Y.~Ling, P.~Liu and M.-H. Wu, \emph{{Learning the black hole metric from
  holographic conductivity}},
  \href{http://dx.doi.org/10.1103/PhysRevD.107.066021}{\emph{Phys. Rev. D} {\bf
  107} (2023) 066021}, [\href{http://arxiv.org/abs/2209.05203}{{\tt
  2209.05203}}].

\bibitem{Jejjala:2023aa}
V.~Jejjala, S.~Mondkar, A.~Mukhopadhyay and R.~Raj, \emph{Learning holographic
  horizons},  \href{http://arxiv.org/abs/2312.08442}{{\tt 2312.08442}}.

\bibitem{Park:2023slm}
C.~Park, S.~Kim and J.~H. Lee, \emph{{Holography Transformer}},
  \href{http://arxiv.org/abs/2311.01724}{{\tt 2311.01724}}.

\bibitem{Carrasquilla_2017}
J.~Carrasquilla and R.~G. Melko, \emph{Machine learning phases of matter},
  \href{http://dx.doi.org/10.1038/nphys4035}{\emph{Nature Physics} {\bf 13}
  (Feb., 2017) 431--434}.

\bibitem{Carrasquilla:2020mas}
J.~Carrasquilla, \emph{{Machine Learning for Quantum Matter}},
  \href{http://dx.doi.org/10.1080/23746149.2020.1797528}{\emph{Adv. Phys. X}
  {\bf 5} (2020) 1797528}, [\href{http://arxiv.org/abs/2003.11040}{{\tt
  2003.11040}}].

\bibitem{Bedolla-Montiel:2020rio}
E.~A. Bedolla-Montiel, L.~C. Padierna and R.~Casta\~neda Priego, \emph{{Machine
  Learning for Condensed Matter Physics}},
  \href{http://dx.doi.org/10.1088/1361-648X/abb895}{\emph{J. Phys. Condens.
  Matter} {\bf 33} (2021) 053001}, [\href{http://arxiv.org/abs/2005.14228}{{\tt
  2005.14228}}].

\bibitem{Suwa:2018twu}
H.~Suwa, J.~S. Smith, N.~Lubbers, C.~D. Batista, G.-W. Chern and K.~Barros,
  \emph{{Machine learning for molecular dynamics with strongly correlated
  electrons}}, \href{http://dx.doi.org/10.1103/PhysRevB.99.161107}{\emph{Phys.
  Rev. B} {\bf 99} (2019) 161107}, [\href{http://arxiv.org/abs/1811.01914}{{\tt
  1811.01914}}].

\bibitem{Boehnlein:2021eym}
A.~Boehnlein et~al., \emph{{Colloquium: Machine learning in nuclear physics}},
  \href{http://dx.doi.org/10.1103/RevModPhys.94.031003}{\emph{Rev. Mod. Phys.}
  {\bf 94} (2022) 031003}, [\href{http://arxiv.org/abs/2112.02309}{{\tt
  2112.02309}}].

\bibitem{Chen:2021giw}
S.~Y. Chen, H.~T. Ding, F.~Y. Liu, G.~Papp and C.~B. Yang, \emph{{Machine
  learning spectral functions in lattice QCD}},
  \href{http://arxiv.org/abs/2110.13521}{{\tt 2110.13521}}.

\bibitem{Zhou:2023pti}
K.~Zhou, L.~Wang, L.-G. Pang and S.~Shi, \emph{{Exploring QCD matter in extreme
  conditions with Machine Learning}},
  \href{http://arxiv.org/abs/2303.15136}{{\tt 2303.15136}}.

\bibitem{Gubser:1998bc}
S.~S. Gubser, I.~R. Klebanov and A.~M. Polyakov, \emph{{Gauge theory
  correlators from non-critical string theory}},
  \href{http://dx.doi.org/10.1016/S0370-2693(98)00377-3}{\emph{Phys. Lett.}
  {\bf B428} (1998) 105--114}, [\href{http://arxiv.org/abs/hep-th/9802109}{{\tt
  hep-th/9802109}}].

\bibitem{Witten:1998qj}
E.~Witten, \emph{{Anti-de Sitter space and holography}}, {\emph{Adv. Theor.
  Math. Phys.} {\bf 2} (1998) 253--291},
  [\href{http://arxiv.org/abs/hep-th/9802150}{{\tt hep-th/9802150}}].

\bibitem{E:2017}
W.~E, \emph{A proposal on machine learning via dynamical systems},
  \href{http://dx.doi.org/10.1007/s40304-017-0103-z}{\emph{Communications in
  Mathematics and Statistics} {\bf 5} (02, 2017) 1--11}.

\bibitem{Sander:2022aa}
M.~E. Sander, P.~Ablin and G.~Peyr{\'e}, \emph{Do residual neural networks
  discretize neural ordinary differential equations?},
  \href{http://arxiv.org/abs/2205.14612}{{\tt 2205.14612}}.

\bibitem{Scheffler_2005}
M.~Scheffler, M.~Dressel, M.~Jourdan and H.~Adrian, \emph{Extremely slow drude
  relaxation of correlated electrons},
  \href{http://dx.doi.org/10.1038/nature04232}{\emph{Nature} {\bf 438} (Dec.,
  2005) 1135--1137}.

\bibitem{Kim:2014bza}
K.-Y. Kim, K.~K. Kim, Y.~Seo and S.-J. Sin, \emph{{Coherent/incoherent metal
  transition in a holographic model}},
  \href{http://dx.doi.org/10.1007/JHEP12(2014)170}{\emph{JHEP} {\bf 12} (2014)
  170}, [\href{http://arxiv.org/abs/1409.8346}{{\tt 1409.8346}}].

\bibitem{Liu_2012}
H.~Liu, \emph{From black holes to strange metals},
  \href{http://dx.doi.org/10.1063/pt.3.1616}{\emph{Physics Today} {\bf 65}
  (jun, 2012) 68--69}.

\bibitem{Faulkner:2010aa}
T.~Faulkner, N.~Iqbal, H.~Liu, J.~McGreevy and D.~Vegh, \emph{From black holes
  to strange metals}, {\emph{MIT-CTP/4105} (03, 2010) },
  [\href{http://arxiv.org/abs/1003.1728}{{\tt 1003.1728}}].

\bibitem{Hartnoll:2008kx}
S.~A. Hartnoll, C.~P. Herzog and G.~T. Horowitz, \emph{{Holographic
  Superconductors}},
  \href{http://dx.doi.org/10.1088/1126-6708/2008/12/015}{\emph{JHEP} {\bf 0812}
  (2008) 015}, [\href{http://arxiv.org/abs/0810.1563}{{\tt 0810.1563}}].

\bibitem{Hartnoll:2008vx}
S.~A. Hartnoll, C.~P. Herzog and G.~T. Horowitz, \emph{{Building a Holographic
  Superconductor}},
  \href{http://dx.doi.org/10.1103/PhysRevLett.101.031601}{\emph{Phys.Rev.Lett.}
  {\bf 101} (2008) 031601}, [\href{http://arxiv.org/abs/0803.3295}{{\tt
  0803.3295}}].

\bibitem{Cubrovic:2009ye}
M.~Cubrovic, J.~Zaanen and K.~Schalm, \emph{{String Theory, Quantum Phase
  Transitions and the Emergent Fermi-Liquid}},
  \href{http://dx.doi.org/10.1126/science.1174962}{\emph{Science} {\bf 325}
  (2009) 439--444}, [\href{http://arxiv.org/abs/0904.1993}{{\tt 0904.1993}}].

\bibitem{Faulkner:2010zz}
T.~Faulkner, N.~Iqbal, H.~Liu, J.~McGreevy and D.~Vegh, \emph{{Strange metal
  transport realized by gauge/gravity duality}},
  \href{http://dx.doi.org/10.1126/science.1189134}{\emph{Science} {\bf 329}
  (2010) 1043--1047}.

\bibitem{Lee:2008xf}
S.-S. Lee, \emph{{A Non-Fermi Liquid from a Charged Black Hole: A Critical
  Fermi Ball}},
  \href{http://dx.doi.org/10.1103/PhysRevD.79.086006}{\emph{Phys.Rev.} {\bf
  D79} (2009) 086006}, [\href{http://arxiv.org/abs/0809.3402}{{\tt
  0809.3402}}].

\bibitem{Liu:2009dm}
H.~Liu, J.~McGreevy and D.~Vegh, \emph{{Non-Fermi liquids from holography}},
  \href{http://dx.doi.org/10.1103/PhysRevD.83.065029}{\emph{Phys. Rev.} {\bf
  D83} (2011) 065029}, [\href{http://arxiv.org/abs/0903.2477}{{\tt
  0903.2477}}].

\bibitem{Faulkner:2009wj}
T.~Faulkner, H.~Liu, J.~McGreevy and D.~Vegh, \emph{{Emergent quantum
  criticality, Fermi surfaces, and AdS(2)}},
  \href{http://dx.doi.org/10.1103/PhysRevD.83.125002}{\emph{Phys.Rev.} {\bf
  D83} (2011) 125002}, [\href{http://arxiv.org/abs/0907.2694}{{\tt
  0907.2694}}].

\end{thebibliography}
\bibliographystyle{JHEP}

\providecommand{\href}[2]{#2}\begingroup\raggedright\endgroup

\end{document}